\documentclass[10pt,journal,compsoc]{IEEEtran}
\usepackage{mathptmx}
\usepackage{graphicx}
\usepackage{times}
\usepackage{epsfig}
\usepackage{amsmath}
\usepackage{graphicx}
\usepackage{amssymb}
\usepackage{subfigure}
\usepackage{amsthm}
\usepackage{array}
\usepackage[linesnumbered]{algorithm2e}
\usepackage{booktabs} 
\usepackage{wrapfig}
\usepackage{color}

\newtheorem{theorem}{Theorem}[section]
\newtheorem{corollary}{Corollary}[theorem]
\newtheorem{lemma}[theorem]{Lemma}
\theoremstyle{definition}
\newtheorem{definition}{Definition}[section]
\theoremstyle{remark}

\usepackage{multirow}

\usepackage{array}
\usepackage{ragged2e}

\ifCLASSOPTIONcompsoc
  \usepackage[nocompress]{cite}
\else
  \usepackage{cite}
\fi
\ifCLASSINFOpdf

\else

\fi
\begin{document}
\title{Top-Down Shape Abstraction Based on Greedy Pole Selection}
\author{Zhiyang~Dou,
Shiqing~Xin$^\star$,
Rui~Xu,
Jian~Xu,
Yuanfeng~Zhou,\\
Shuangmin~Chen,
Wenping~Wang,
Xiuyang~Zhao,
Changhe~Tu
\IEEEcompsocitemizethanks{   
\IEEEcompsocthanksitem Zhiyang Dou, Shiqing Xin, Rui Xu, Yuanfeng Zhou and Changhe Tu are with Shandong University, China. Jian Xu is with Ningbo Institute of Materials Technology and Engineering, Chinese Academy of Sciences, Ningbo, China. Shuangmin Chen is with Qingdao University of Science and Technology, China. Wenping Wang is with The University of Hong Kong. Xiuyang Zhao is with  University of Jinan, China.\protect\\
Correspondence to Shiqing Xin at xinshiqing@163.com.
}
\thanks{Manuscript received May 15, 2020; revised May 15, 2020.}}
\markboth{IEEE TRANSACTIONS ON VISUALIZATION AND COMPUTER GRAPHICS, VOL. XX, NO. X, MONTH YEAR}%
{Shell \MakeLowercase{\textit{et al.}}: Bare Advanced Demo of IEEEtran.cls for IEEE Computer Society Journals}

\IEEEtitleabstractindextext{%
\begin{abstract}
\justifying
Motivated by the fact that the medial axis transform is able to encode the shape completely, we propose to use as few medial balls as possible to approximate the original enclosed volume by the boundary surface. We progressively select new medial balls, in a top-down style, to enlarge the region spanned by the existing medial balls. The key spirit of the selection strategy is to encourage large medial balls while imposing given geometric constraints. We further propose a speedup technique based on a provable observation that the intersection of medial balls implies the adjacency of power cells (in the sense of the power crust).\\
We further elaborate the selection rules in combination with two closely related applications. One application is to develop an easy-to-use ball-stick modeling system that helps non-professional users to quickly build a shape with only balls and wires, but any penetration between two medial balls must be suppressed. The other application is to generate porous structures with convex, compact (with a high isoperimetric quotient) and shape-aware pores where two adjacent spherical pores may have penetration as long as the mechanical rigidity can be well preserved.
\end{abstract}

}
\maketitle
\IEEEdisplaynontitleabstractindextext
\IEEEpeerreviewmaketitle

\ifCLASSOPTIONcompsoc
\IEEEraisesectionheading{\section{Introduction}\label{sec:introduction}}
\else
\section{Introduction}
\label{sec:introduction}
\fi
\IEEEPARstart{S}{hape} abstraction draws much attention in computer vision and pattern recognition since it is closely related to shape recognition and shape understanding. One of the key motivations of shape abstraction is parsimony of description, i.e., an object could be
described by relatively few primitives, each of
which in turn requiring only a few parameters~\cite{tulsiani2017learning}. It is a hard yet fascinating problem~\cite{borges1997class}.

Different from explaining objects with volumetric primitives based on learning techniques~\cite{tulsiani2017learning}, in this paper, we study the problem of approximating the original enclosed volume using as few medial balls as possible motivated by the fact that the medial axis transform is able to encode nearly the complete shape, which approaches this problem motivated by the geometry.

It's well known that the full set of medial balls defines the enclosed volume $\Omega$, and thus we intend to use a subset of representative medial balls to generate the shape abstraction of $\Omega$. In the discrete setting, the medial surface can be approximated by the Voronoi diagram w.r.t.~a set of samples extracted from the boundary surface, while the medial balls can be replaced by polar balls centered at a subset of Voronoi vertices (also named as {\em poles})~\cite{amenta1998new,amenta2001power:A,amenta2001power:B}. We propose to progressively select new polar balls, in a top-down style, to enlarge the region occupied by the existing polar balls. The key point of selecting poles is to encourage large polar balls subject to given geometric constraints.

In implementation, we require an efficient proximity query technique to determine how a given polar ball intersects with a set of existing polar balls. Interestingly, we observe that there is a provably effective speedup technique that the intersection of polar balls implies the adjacency of power cells (in the sense of the power crust~\cite{amenta2001power:A,amenta2001power:B,miklos2010discrete,attali1996modeling,dey2004approximating,giesen2006medial}). Therefore, we maintain a dynamic power diagram w.r.t. a set of poles to support efficient proximity query rather than detect the intersection between polar balls in a brute-force manner.
We further put forward two closely related applications including ball-stick modeling and porous structure generation.

\begin{figure*} 
\centering
 \includegraphics[width=17cm]{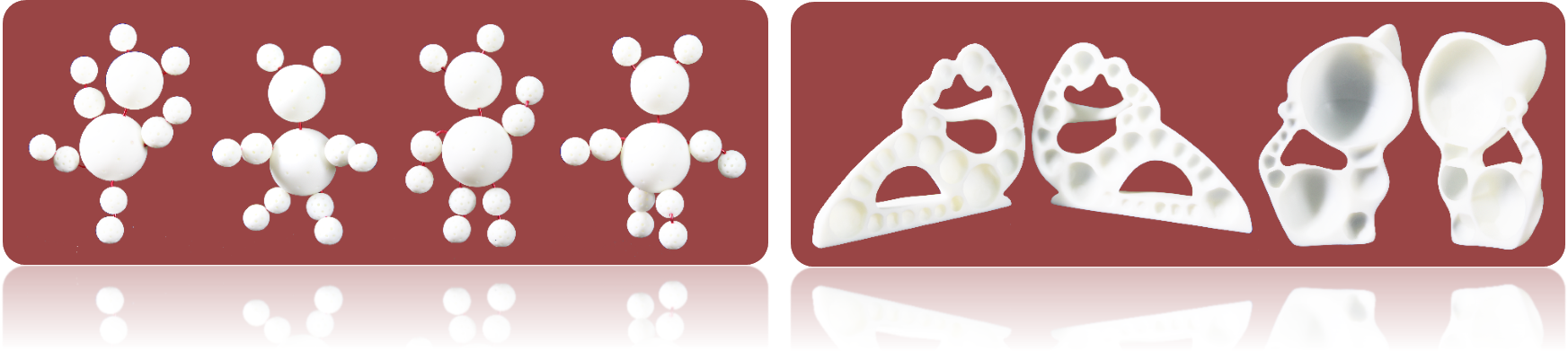}
  \caption{Medial ball based shape abstraction. (a)~Ball-stick toy design where the Teddy-Bear model is abstractly represented by a set of balls (with four customized sizes) and wires so as to be reshaped to various poses. (b)~Cost-effective porous structure generation where adjacent spherical pores may have penetration but every pore has a compact shape for purpose of preserving rigidity.\label{fig:teaser}}
\end{figure*}

Our contributions are three-fold:
\begin{enumerate}
\vspace{-1mm}
  \item We propose a greedy selection strategy that approximates the enclosed volume with as few as possible polar balls.
  \item We prove that the intersection of polar balls implies the adjacency of power cells, which leads to an effective speedup technique; See Theorem~\ref{thm:interference} in Section~\ref{sec:background}.
  \item We elaborate the selection rules in combination with two closely related applications and validate the uses and effectiveness of our algorithm, See Figure~\ref{fig:teaser}.
\end{enumerate}

\vspace{-2mm}

\section{Related Work}
\textbf{Medial surface.}
The medial axis transform~(MAT), as a complete shape descriptor~\cite{blum1967transformation}, is central to various applications such as shape recognition/manipulation and surface approximation~\cite{sherbrooke1995computation}. Given a closed, oriented and bounded 2-manifold surface $\mathcal{S}$ in $\mathbb{R}^3$, a ball $B$ inside the volume $\Omega$ enclosed by $\mathcal{S}$ is said to be {\em maximal} if no other ball in $\Omega$ contains $B$. In fact, the medial surface is formed by the locii of the centers of maximal balls inside $\Omega$. Furthermore, the union of maximal balls defines $\Omega$.

Generally the MAT problem doesn't have a closed-form solution, and one has to seek for a numerical solution instead.
The most popular technique is to initialize the medial surface using the Voronoi diagrams w.r.t.~a set of sufficiently dense boundary samples and then prune spikes or optimize mesh tessellations based on various rules. There are many existing approaches on this side, such as $\lambda$-medial axis~\cite{chazal2005lambda}, quadratic error metric (QEM) based MAT~\cite{li2015q}, delta medial axis~(DMA)~\cite{marie2016delta}, bending potential ratio (BPR) pruning~\cite{shen2011skeleton}, erosion thickness (ET) measure~\cite{yan2016erosion} and voxelization based $\lambda$ pruning~\cite{yan2018voxel}. The simplified MAT, without doubt, cannot exactly recover the original boundary surface, and thus one has to control the error metric by computing the Hausdorff distance~\cite{sun2013medial} or adaptively classifying geometric features~\cite{miklos2010discrete}.

The power crust~\cite{amenta2001power:A,amenta2001power:B,miklos2010discrete,attali1996modeling,dey2004approximating,giesen2006medial} also begins with Voronoi diagrams w.r.t. boundary samples, but uses a subset of Voronoi vertices as poles to generate MAT approximation by computing a power diagram w.r.t.~poles. In fact, it builds an inverse transform from the MAT to the surface representation, and thus facilitates many applications such as surface reconstruction and surface offsetting.
The theme of this paper, i.e. greedily selecting some poles for shape abstraction, is closely related to the power crust theory. We shall give the theoretical foundation in the next section.

\noindent \textbf{Shape abstraction.}
Shape abstraction is an important research topic in computer vision and pattern recognition and central to shape recognition and shape understanding~\cite{tulsiani2017learning}. On some occasions it is also called shape averaging or shape simplification.
Abstract shape representations may vary with different contexts. Curves~\cite{mehra2009abstraction}, contours~\cite{duta1999learning,mehra2009abstraction}, skeletons~\cite{au2008skeleton} and even black-and-white images~\cite{lin2018scale} are all commonly used abstract shape representation forms. There is a large body of literature on how to approximate the given surface/volume using a set of shape primitives that include planar faces~\cite{mccrae2013surface,cohen2004variational}, cylinders, spheres~\cite{thiery2013sphere}, tori~\cite{kaiser2018survey}, or even superquadrics~\cite{barr1981superquadrics}, hyperquadrics~\cite{hanson1988hyperquadrics} and blobby models~\cite{muraki1991volumetric}.
An important application of shape abstraction is to help understand the 3D structure/layout of a given scene, as well as the spatial, functional, and semantic relationships between objects in the scene~\cite{mitra2013structure,biasotti2016recent}.
Beside the above mentioned geometry based approaches, Gestalt principles can also serve as a guide to shape simplification and abstraction~\cite{lovset2013rule, dang2014safe}, especially useful for automatically simplifying line drawings of architectural buildings~\cite{nan2011conjoining,Kratt2018-09Sketc-41240}.

In this paper, we select as few medial balls as possible to approximate the original enclosed volume by the boundary surface, which is from a geometric perspective, rather than based on learning techniques~\cite{tulsiani2017learning}.  We define a priority measure to encourage large medial balls while suppressing serious penetration between medial balls.

\noindent \textbf{Easy modeling.} Simple modeling tools are becoming popular nowadays - nonprofessional users can build their own favorite toys/products with a 3D printer. For example, Igarashi et al.~\cite{igarashi2007teddy} proposed an interactive system that supports constructing a 3D polygonal
surface from the 2D silhouette. Mori et al.~\cite{mori2007plushie} introduced the physical simulation technique to help design original plush toys. They also introduced simple physical simulation to improve digital fabrication.
Nealen et al.~\cite{nealen2007fibermesh} proposed to design freeform surfaces with a collection of 3D curves. Users can not only use strokes as handles to control the geometry, but also add, remove, and deform these control curves easily.
Igarashi et al.~\cite{igarashi2012beady} introduced an interactive system named ``Beady'' to assist the design and construction of customized 3D beadwork. Attene et al.~\cite{attene2006hierarchical} proposed to approximate a 3D model with a set of planes, spheres and cylinders based on hierarchical face clustering, but they allow intersection between shape primitives.
Besides, various easy modeling approaches are applied in education, art production, products design, and pattern recognition. Generally speaking, the convenience of easy modeling comes from at least two aspects: (1)~an image, or a point cloud, or a 3D digital model as the hint and (2)~intelligent/fast computation to facilitate real-time user interaction.

In fact, easy modeling with a set of penetration-free shape primitives of prescribed types is our main interest.  One of our applications in this paper is to develop an easy-to-use ball-stick modeling system that helps children to quickly build a shape with only balls and wires. The key lies in how to approximately represent the enclosed volume by a set of penetration-free medial balls with various sizes (typically customized in advance).

\noindent \textbf{Porous structure.}
Porous structures are ubiquitous in nature and widely used in our daily life because of nice physical properties.
For example, porous solids have been proved to be good candidates as the carbon dioxide recycling sorbents~\cite{Lu2013Porous,jakus20183d}.
Porous structure design has drawn much attention in recent years especially with the innovation of the additive manufacturing technology. An important research topic is to reduce the material cost, as well as the weight, of a given object while providing a durable printed model that is resistant to impact and external forces~\cite{lu2014build}. Wang et al.~\cite{wang2013cost} discussed cost-effective 3D printing by reinforcing a thin shell with skin-frame structures.
Lu et al.~\cite{lu2014build} proposed a hollowing optimization algorithm based on the concept of honeycomb structure.
Recently, Mao et al.~\cite{mao2018generating} used hybrid structures for designing the support structure.

In fact, finite element analysis of stress is central to fabrication oriented design.
For example, Stava et al.~\cite{stava2012stress} suggested strengthening printed objects by hollowing, local thickening, and adding extra struts. There is a trend of addressing these issues based on topology optimization~\cite{liu2018current,bendsoe2001topology,wang2003level,allaire2004structural}.
Wu et al. ~\cite{wu2016self} proposed a method to generate application-specific infill structures on rhombic cells so that the resultant structures can automatically satisfy manufacturing requirements on overhang-angle and wall-thickness.
Groen et al.~\cite{groen2018optimization} introduced an efficient homogenization based approach to perform topology optimization of coated structures with orthotropic infill material. We refer to~\cite{liu2018current} for a survey.

Inspired by the structure of the spherical inverse opal, we notice that the convexity and compactness (high isoperimetric quotient) of pores are central to preserving mechanical rigidity. Therefore, in this paper, we proposed to use medial balls as primitives to generate pores but allow moderate penetration between neighboring pores (we add an additional partition board to separate neighboring pores). In implementation, our algorithm is based on a greedy selection strategy that gives a higher priority to those large medial balls but suppresses serious penetration between medial balls.

\section{Theoretical Background}\label{sec:background}

\subsection{Voronoi diagram \& power diagram}
\begin{figure}[!ht]
    \centering
       \includegraphics[width=0.78\linewidth]{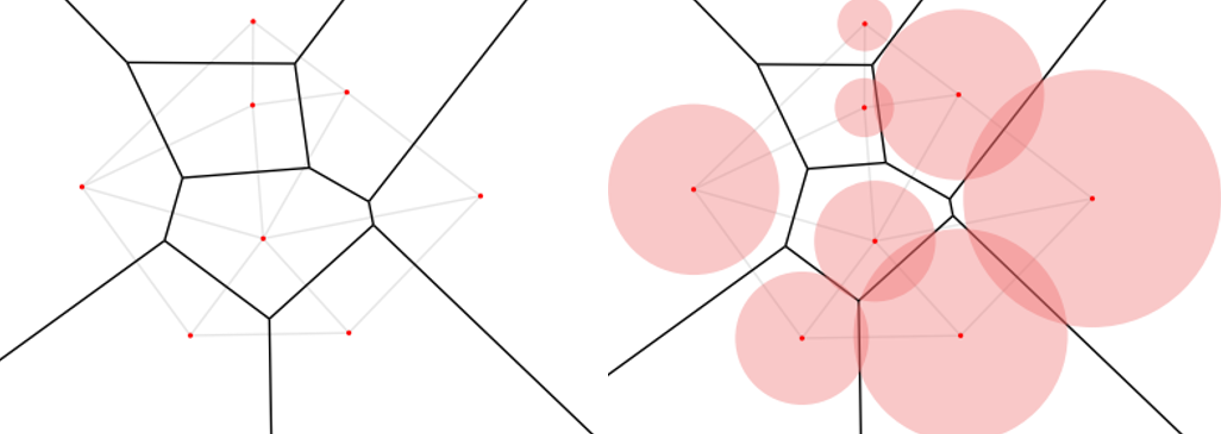}
    \makebox[0.2\linewidth][r]{Voronoi diagram}\makebox[0.36\linewidth][r]{Power diagram}
    \caption{Voronoi diagram and power diagram in 2D.}\label{fig:Voronoi:power}
\end{figure}
    \vspace{-2mm}
Voronoi diagrams are used to partition a given domain $\Omega\subset \mathbb{R}^d$ into regions based on the straight-line distance to a set of generators $\{\mathbf{x}_i\in\Omega\}_{i=1}^n$ such that $\mathbf{x}_i$ dominates the subregion (also called a cell):
\begin{equation}
\Omega_i^{\text{vor}}:~\{\mathbf{x}\in\Omega\;\big|\;\|\mathbf{x}-\mathbf{x}_i\|\leq\|\mathbf{x}-\mathbf{x}_j\|,j\neq i\}.
\end{equation}
Power diagrams can be viewed as an extension of Voronoi diagrams, where each generator $\mathbf{x}_i$
is equipped with a weight $w_i$ to adjust the influence range. That is to say, by assuming the power distance $d^{\text{pow}}(x,x_i)$ between $x$ and the weighted generator $x_i$ to be $\|\mathbf{x}-\mathbf{x}_i\|^2-w_i$, the cell dominated by $\mathbf{x}_i$ becomes
\begin{equation}
\Omega_i^{\text{pow}}:~\{\mathbf{x}\in\Omega\;\big|\;d^{\text{pow}}(x,x_i)\leq d^{\text{pow}}(x,x_j),j\neq i\}.
\end{equation}
A generator with a larger weight is more dominant, and when all the weights are equal, the power diagram reduces to a Voronoi diagram. Different from the Voronoi diagram, there may be some generators hidden from the power diagram, which means these generators have no corresponding dominant regions. In Figure~\ref{fig:Voronoi:power}, we show a Voronoi diagram and a power diagram respectively.

\subsection{Medial surface}
\begin{wrapfigure}{r}{2.1cm}
  \vspace{0mm}
  \hspace*{-2mm}
  \centerline{
  \includegraphics[width=2.4cm]{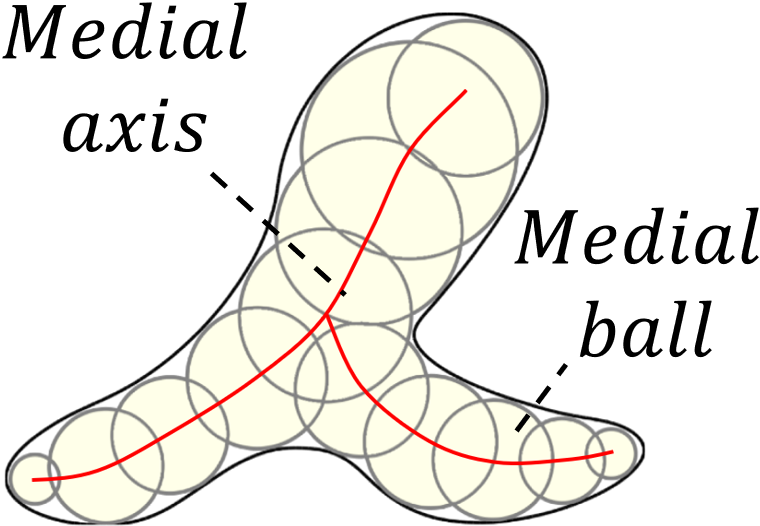}
  }
  \vspace*{0mm}
\end{wrapfigure}
Suppose that $\mathcal{S}$ is a closed, oriented and bounded 2-manifold surface in $\mathbb{R}^3$. By the term {\em medial surface}, we mean the loci given by the center of a moveable sphere inside $\mathcal{S}$ that has at least two touching points with $\mathcal{S}$. The medial surface $\mathcal{M}$ of $\mathcal{S}$, combined with the medial radius function $\mathcal{R}$, is called {\em medial-axis transform} (MAT), denoted by a pair $\big(\mathcal{M}, \mathcal{R}\big)$.

\subsection{Feature-preserving sampling}
\begin{definition}\label{dfn:lfs}
Given a surface $\mathcal{S}$ and a point $s\in\mathcal{S}$, the {\em local feature size}~(LFS) of $s$, denoted by ${LFS}(s)$, is defined to be the distance between $s$ and the medial surface $\mathcal{M}$.
\end{definition}

\begin{definition}\label{dfn:sampling}
Let $\{s_i\}_{i=1}^n$ be a set of samples extracted from $\mathcal{S}$. If $\forall s\in\mathcal{S}$, there exists a sample $s_i$ such that $\|s-s_i\|\leq \epsilon$, then $\{s_i\}_{i=1}^n$ is an $\epsilon$-dense sample set. If $\forall s\in\mathcal{S}$, there exists a sample $s_i$ such that $\|s-s_i\|\leq \alpha\times{LFS}(s)$, then $\{s_i\}_{i=1}^n$ is an $\alpha$-LFS sample set.
\end{definition}

\subsection{Inside/outside poles}
\begin{wrapfigure}{r}{2cm}
  \vspace*{-2mm}
  \hspace*{-3.4mm}
  \centerline{
  \includegraphics[width=2.2cm]{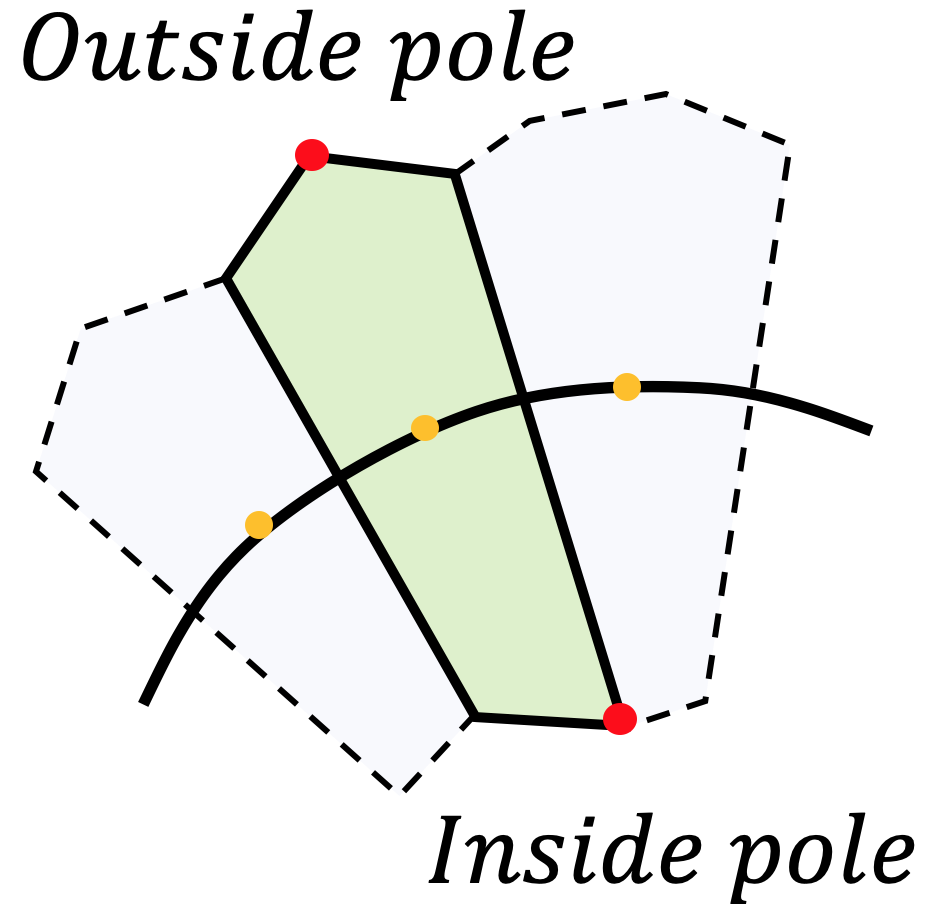}
  }
  \vspace*{-5mm}
\end{wrapfigure}
In the discrete setting, the Voronoi diagrams $\mathcal{V}$ w.r.t.~a set of sufficiently dense boundary samples in $\mathcal{S}$ are commonly used to initialize the medial surface.
Given an $\alpha$-LFS sample set, each sample $s\in\mathcal{S}$ dominates a cell across $\mathcal{S}$ (generally a sufficiently large box is used to make $s$'s cell to be closed). On either side of the surface, there is a cell vertex farthest from $s$, respectively named an {\em inside pole} and an {\em outside pole}; See the inset figure.
Each pole $p$ defines a {\em polar ball} $B(p,\|p-s\|)$, where $s$ is the sample point closest to $p$.
\begin{definition}\label{dfn:power crust}
The {\em power crust} is the boundary between the power diagram cells belonging to inside poles and power diagram cells belonging to outside poles.
\end{definition}
The power crust is able to approximate the original boundary surface under the condition that the samples are sufficiently dense~\cite{amenta2001power:B}.
\begin{lemma}\label{lem:crust}
Let $P=\{p_i\}$ and $Q=\{q_i\}$ be respectively the inside and outside poles w.r.t.~an $\alpha$-LFS sample set. When $\alpha$ approaches 0, the power crust given by $P$ and $Q$ approximates the surface~$\mathcal{S}$.
\end{lemma}
It can be shown that the following observation naturally holds based on Lemma~\ref{lem:crust}. (More theoretical results are available in~\cite{amenta2001power:B,choi1997mathematical,choi2001hyperbolic},
and we organize them in Appendix~A.)
\begin{corollary}\label{cor:volume-aprox}
Let $P=\{p_i\}$ be the inside pole set w.r.t.~an $\alpha$-LFS sample set.
When $\alpha$ approaches 0, the union of the polar balls $$\{B(p_i,r_i)\;\vert\; p_i~\text{~is~an~inside~pole}\}$$ approximates the volume enclosed by $\mathcal{S}$.
\end{corollary}
\noindent
$\mathbf{Remark.}$~Due to the approximation ability of polar balls, we make no distinction between polar balls and medial balls in later sections.
\subsection{Proximity query between polar balls}
Corollary~\ref{cor:volume-aprox} motivates us to use a subset of representative inside poles $\{p_i\}_{i=1}^K$, or alternatively their corresponding polar balls $\{B(p_i,r_i)\}_{i=1}^K$, to generate the shape abstraction of the whole enclosed volume. For this purpose, we adopt a top-down pole selection strategy that encourages large polar balls while suppressing serious intersection between polar balls.

In implementation, we require an efficient proximity query technique to determine how a given polar ball intersects with a set of existing polar balls. In fact, there is a fundamental relationship between the proximity of polar balls and the proximity of power cells.
\begin{lemma}\label{lem:equal-energy}
Suppose that $B(p_1,r_1)$ and  $B(p_2,r_2)$ are two balls in $\mathbb{R}^3$. If $B(p_1,r_1)$ intersects the ball $B(p_2,r_2)$ at a disk, for every point $q$ in the plane $\pi$ of the disk, $q$ has equal weighted distances to $p_1$ and $p_2$, i.e. $d^{\text{pow}}(q,p_1)=d^{\text{pow}}(q,p_2),$ where the weight is set to be the squared radius.
\end{lemma}
In other words, under the condition that the polar ball $B(p_1,r_1)$ intersects the ball $B(p_2,r_2)$ at a disk, the disk must be coplanar with the {\em equal-power} plane w.r.t.~$p_1$ and $p_2$ (each pole is equipped with a weight of the squared medial radius).
We further observe that the power diagram is able to serve as an effective tool for polar-ball based dynamic proximity query, which is the following theorem.
\begin{theorem}\label{thm:interference}
Let $P=\{p_i\}$ and $Q=\{q_i\}$ be respectively the inside poles and the outside poles w.r.t.~a sufficiently dense $\alpha$-LFS sample set on the surface $\mathcal{S}$. Suppose that we have a subset of inside polar balls centered at
$P^+\subset P$. For another inside pole $p^\text{new}\in P\setminus P^+$, the statement that the polar ball centered at $p^\text{new}$ intersects with one of the existing polar balls $\{B(p_i,r_i):p_i\in P^+\}$ implies that there exists $p_j\in P^+$ such that $p^\text{new}$'s cell is neighboring to $p_j$'s cell in the power diagram w.r.t.~$P^+\bigcup Q\bigcup p^\text{new}$.
\end{theorem}

\begin{figure}[h]
    \centering
    \includegraphics[width=0.45\linewidth]{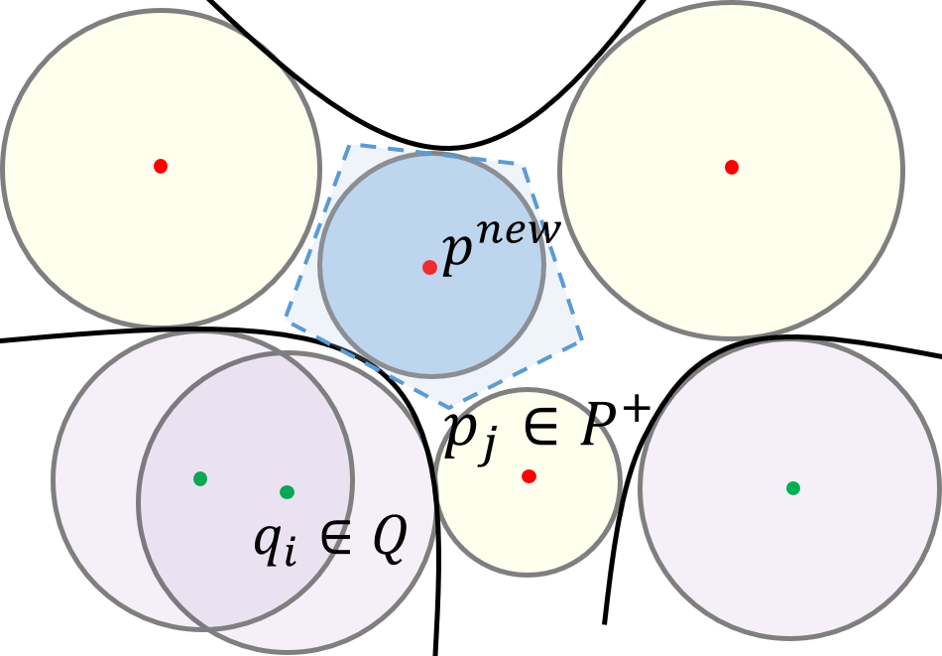}\quad
    \includegraphics[width=0.45\linewidth]{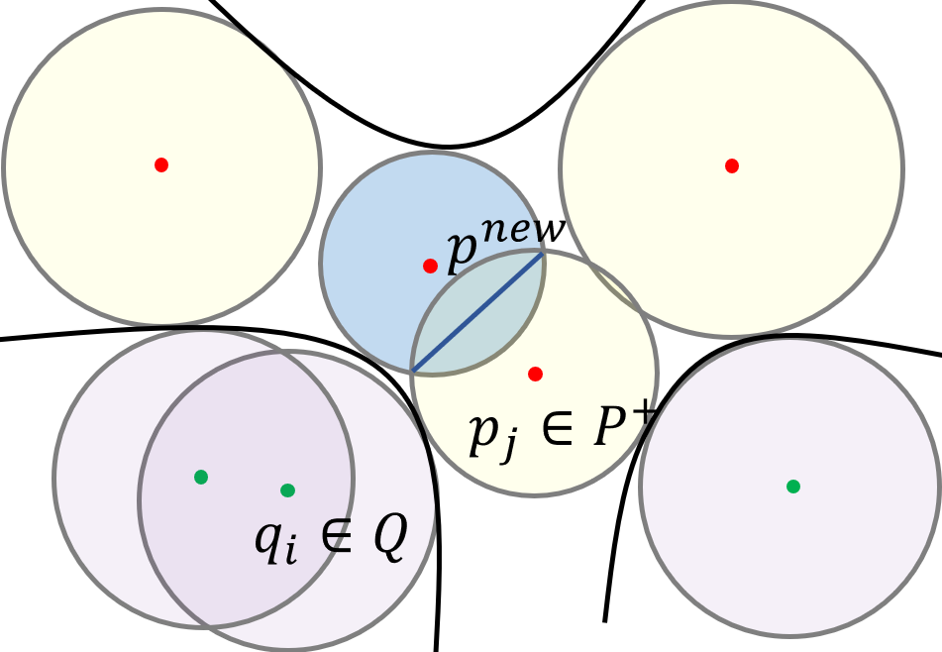}\\
    \makebox[0.49\linewidth][c]{(a)}\makebox[0.49\linewidth][c]{(b)}\\
    \vspace{1mm}
    \caption{Proof to Theorem~\ref{thm:interference}. (a)~If $p^\text{new}$'s cell is not neighboring to any cell given by $P^+\bigcup Q$, then the polar ball centered at $p^\text{new}$ is totally inside $p^\text{new}$'s cell. (b)~If $p^\text{new}$'s medial ball intersects $p_j$'s medial ball, then the equal-power plane between them passes through $p^\text{new}$'s cell.}
    \label{fig:appendix_prove}
\end{figure}

\begin{proof}
We consider the power diagram w.r.t.~$P^+\bigcup Q\bigcup p^\text{new},$ where each pole is weighed by the square of the radius of its polar ball.

Suppose that the pole $p^\text{new}$ is defined based on some boundary sample point $s$, which implies that $p^\text{new}$ is closer, or at least equally distant, to $s$ than any other poles belonging to $P^+\bigcup Q$. It is easy to verify that  $d^\text{pow}(s, p^\text{new})=0$. Considering that $s$ is located outside/on the polar balls given by $P^+\bigcup Q$, the power distance between $s$ and any pole in $P^+\bigcup Q$ is non-negative. Therefore, $p^\text{new}$'s cell is not empty (since the cell contains at least $s$). It can be further concluded that $p^\text{new}$'s cell is convex.

Now we assume that $p^\text{new}$'s cell is not neighboring to any cell given by $P^+\bigcup Q$, which implies that the boundary of $p^\text{new}$'s cell is formed by $p^\text{new}$ and the poles in $Q$. On one hand, we can see that  $d^\text{pow}(t, p^\text{new})\geq 0$ for any point $t$ on the boundary of $p^\text{new}$'s cell (since $t$ has equal power distances to $p^\text{new}$ and some outside pole $q_i$ in $Q$; but the power distance between $t$ and any outside pole in $Q$ is non-negative), which implies that the polar ball centered at $p^\text{new}$ is totally inside $p^\text{new}$'s cell (note that $d^\text{pow}(t, p^\text{new})=0$ for any point $t$ on the surface of the polar ball centered at $p^\text{new}$); See Figure~\ref{fig:appendix_prove}(a). On the other hand, the equal-power plane between $p^\text{new}$ and any pole in $P^+$ cannot pass through $p^\text{new}$'s cell since $P^+$ doesn't contribute to the boundary of $p^\text{new}$'s cell at all.

According to the given condition that $p^\text{new}$'s polar ball intersects with one of the polar balls given by $P^+$, without loss of generality, we assume that  $p^\text{new}$'s polar ball intersects with $p_j$'s polar ball, where $p_j\in P^+$. The two polar balls intersects at a disk and any point in this disk is located inside $p^\text{new}$'s polar ball and thus also inside $p^\text{new}$'s cell. However, the disk is coplanar with the equal-power plane between $p^\text{new}$ and $p_j$, which shows that this equal-power plane is given by $p^\text{new}$ and a pole in $P^+$ but passes through $p^\text{new}$'s cell (see Figure~\ref{fig:appendix_prove}(b)), which contradicts to the above assumption.
\end{proof}

\begin{corollary}\label{cor:interference}
Let $P=\{p_i\}$ and $Q=\{q_i\}$ be respectively the inside poles and the outside poles w.r.t.~a sufficiently dense $\alpha$-LFS sample set on the surface $\mathcal{S}$. Suppose that we have two inside polar ball sets centered at $P_1,P_2\subset P$ with $P_1\bigcap P_2 \neq \emptyset$. The statement that $P_1$'s polar balls intersect with $P_2$'s polar balls implies that there exists two inside poles $p_1\in P_1$ and $p_2\in P_2$ such that $p_1$'s cell is neighboring to $p_2$'s cell in the power diagram w.r.t.~$P_1\bigcup P_2\bigcup Q$.
\end{corollary}

\section{Problem formulation}

Suppose that we have obtained a full set of inside poles, i.e. $P=\{p_i\}_{i=1}^n$, which define a set of medial-axis balls $\{B(p_i,r_i)\}_{i=1}^n$ inside the volume enclosed by the surface $\mathcal{S}$. Our task is to develop a strategy on how to incrementally select some representative poles, generally just a few, to approximate the volume as far as possible under some specified geometric constraints.

In particular, we shall discuss the pole selection in detail in combination with two separate applications: (1)~ball-stick toy design where we need to use a set of separate medial balls with various sizes, connected by wires, to convey the abstracted shape, and (2)~porous structure generation where two spherical pores may have penetration but every pore must have a compact shape (with a high isoperimetric quotient).
See Section~\ref{sec:ball-stick} and Section~\ref{sec:porous} respectively.

\section{Ball-Stick Toy Building}\label{sec:ball-stick}
\begin{figure}[ht]
    \centering
    \includegraphics[width=0.5\linewidth]{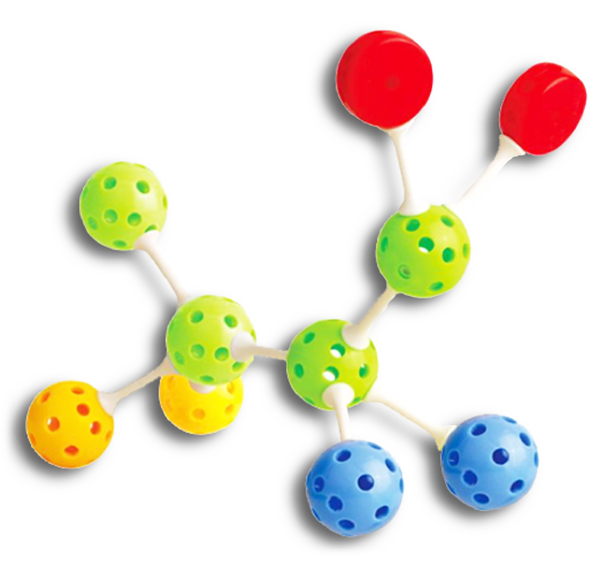}
    \vspace{1mm}
    \caption{A real-life ball-stick toy provided courtesy of Jinchang Toy Factory, Shenzhen City, China.\label{fig:real-toy}}
\end{figure}
We first discuss the shape approximation application in ball-stick toy building.
Many experts in teaching children found that modeling various shapes with some basic components, e.g., clay or magnetic sheets,  have a very positive influence on intelligence development and can make children become more creative. Ball-stick toy design (see Figure~\ref{fig:real-toy}), i.e. assembling balls and sticks (or wires) into a target shape, is one of the most popular educational games.
We find that the technical principles behind the ball-stick toy design have a close relationship to the theme of this paper, i.e. medial ball based shape abstraction.
\subsection{Selection strategy}
\begin{figure}[ht]
    \centering
    \includegraphics[width=0.45\linewidth]{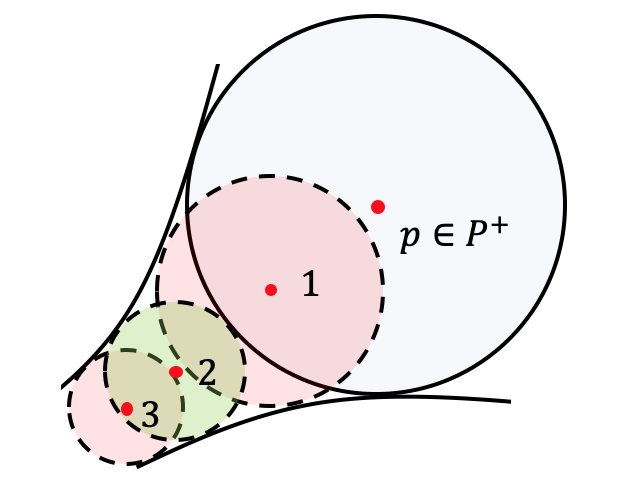}
    \vspace{1mm}
    \caption{Suppose that the medial ball colored in light green is already selected in $P^+$. Considering that No.~1 medial ball has intersection with the existing medial ball whereas No.~3 medial ball is not large enough, we select No.~2 medial ball as the next candidate and append it to $P^+$.\label{ball-stick-selection}}
\end{figure}
Let $P$ be the full set of inside poles.
Technically speaking, we have to select a subset of polar balls, generally not many, centered at a pole set $P^+\subset P$ such that the union of balls $\bigcup\{ B(p_i,r_i):p_i\in P^+\}$ roughly represents the enclosed volume of the input shape. Note that any penetration is not allowed between selected polar balls.

Initially, $P^+$ is an empty set. We have to select the largest polar ball and add the corresponding pole into $P^+$. After that, we enlarge the set $P^+$ by repeatedly selecting a new inside pole $p^\text{new}\in P\setminus P^+\triangleq P^-$.
A natural idea is to use the following greedy strategy to incrementally enlarge $P^+$ until the selected medial balls, after connected, sufficiently resemble the entire shape; See Figure~\ref{ball-stick-selection} for illustration.
\begin{equation}\label{eq:selection}
 p^\text{new}=\mathop{\arg\max}_{p_j\in P^-}\{r_j\::\:|p_j-p_i|-r_i-r_j\geq 0, \forall p_i\in P^+\}.
\end{equation}
We denote $d_{ij}=|p_j-p_i|-r_i-r_j$, and then Eq.~(\ref{eq:selection}) can be rewritten as
\begin{equation}\label{eq:selection2}
 p^\text{new}=\mathop{\arg\max}_{p_j\in P^-}\{r_j\::\:d_{ij}\geq 0, \forall p_i\in P^+\}.
\end{equation}
It's easy to verify the following observation.
\begin{theorem}\label{thm:non-increasing}
  Eq.~(\ref{eq:selection2}) reports a non-increasing medial-radius sequence.
\end{theorem}
Based on Theorem~\ref{thm:non-increasing}, we can stop the selection process until the medial radius is less than a prescribed tolerance~$\delta$. The pseudo-code of a na\"{i}ve implementation is described in Algorithm~\ref{alg:ball-stick-naive}.

\begin{algorithm}[!htb]

\KwIn{The set of inside poles $P$, each associated with a medial radius, and the tolerance of the minimum medial radius~$\delta$ as the termination condition.}
\KwIn{\bf Initialization:} An empty set $P^+$ and an empty priority queue $\mathcal{Q}$. Push all the poles in $P$ to $\mathcal{Q}$ where the medial radius serves as the priority; \\
\Repeat{ $p^\text{new}$'s medial radius is less than $\delta$.}{
        Take the top pole $p^\text{new}$ in $\mathcal{Q}$;\\
        \For{every pole $p$ in $P^+$}
           {
            Check the intersection between $p^\text{new}$'s medial ball and $p$'s medial ball;\\
           }
        \If{there is no intersection}
        {
            Add $p^\text{new}$ into $P^+$;\\
        }
      }

\caption{Na\"{i}ve algorithm for generating penetration-free medial balls.\label{alg:ball-stick-naive}}
\label{alg:one}
\end{algorithm}
\noindent
$\mathbf{Remark.}$~Algorithm~\ref{alg:ball-stick-naive} is not efficient because we have to check the intersection, one by one, between $p^\text{new}$'s medial ball and the existing medial balls given by $P^+$. We shall discuss the speedup strategy int the following subsection.
\subsection{Algorithm speedup}\label{subsec:speedup}
Recall that during each iteration of Algorithm~\ref{alg:ball-stick-naive}, we have to test if the top-priority pole popped from the priority queue is able to define a medial ball that doesn't have penetration with those medial balls given by $P^+$.
The key of speeding up Algorithm~\ref{alg:ball-stick-naive} lies in how to quickly report those candidate poles in $P^+$ that are ``sufficiently close'' to $p^\text{new}$ such that penetration between medial balls may occur, rather than in a brute-force manner.

\begin{figure}[ht]
    \centering
    \includegraphics[width=0.6\linewidth]{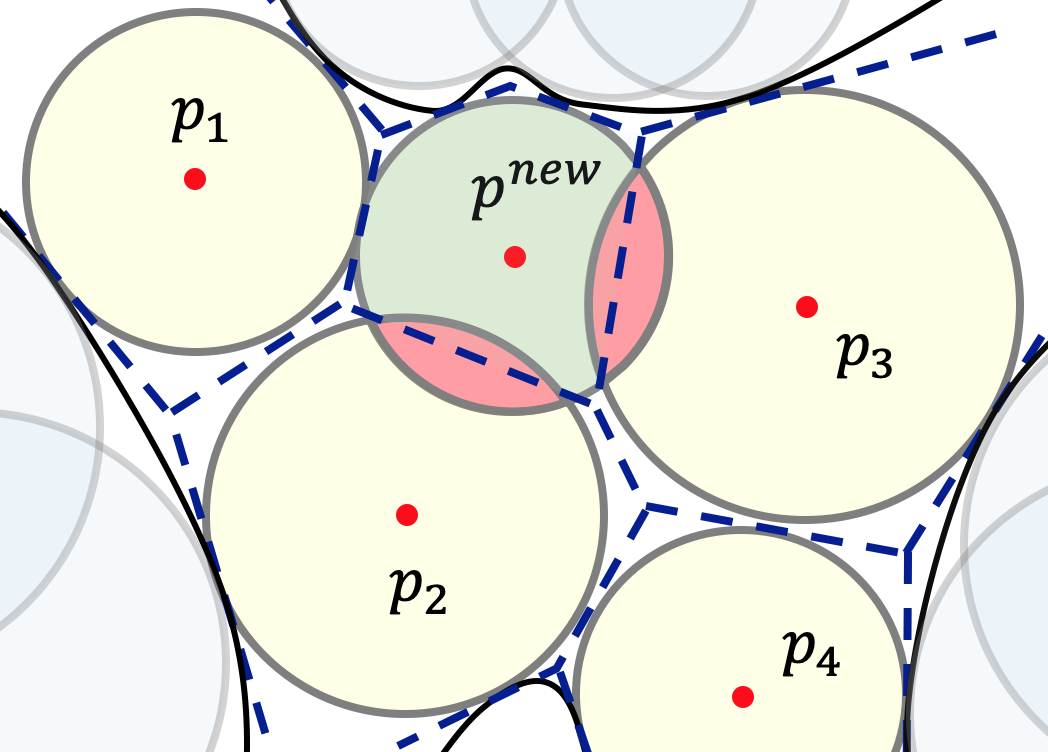}
    \vspace{1mm}
    \caption{Algorithm speedup based on Theorem~\ref{thm:interference}.}
    \label{fig:priority_pic}
    \vspace{-1mm}
\end{figure}

Theorem~\ref{thm:interference} states that the intersection between medial balls implies the adjacency between power cells.
Suppose that we maintain a dynamic power diagram $\mathbf{PD}$ throughout the algorithm. Now we update the power diagram $\mathbf{PD}(P^+\bigcup Q)$ to $\mathbf{PD}(P^+\bigcup Q\bigcup p^\text{new})$ ($Q$ is the set of outside poles), which requires only a little cost due to the incremental change of partition configuration.
As Figure~\ref{fig:priority_pic} shows, the medial balls colored in yellow are defined by $P^+$ while the medial ball colored in green is given by $p^\text{new}$ that has a medial radius $r^\text{new}$. We can see that $p^\text{new}$'s cell is neighboring to 3 existing poles, i.e. $p_1,p_2,p_3$ in $P^+$. Theorem~\ref{thm:interference} tells us $p^\text{new}$'s medial ball cannot intersect with those medial balls given by $P^+\setminus\{p_1,p_2,p_3\}.$ Therefore, it suffices to check the assertions $|p^\text{new}-p_i|-r^\text{new}-r_i\geq 0,i=1,2,3,$ one by one. If all the assertions are true, we accept $p^\text{new}$ as the next candidate and add it into $P^+$; Otherwise, we consider the next top-priority pole popped from the priority queue. Detailed performance contrast between the brute-force method and the accelerated version will be given in Section~\ref{sec:evaluation}.

\subsection{Connection between poles}
\begin{figure}[ht]
    \centering
    \includegraphics[width=0.36\linewidth]{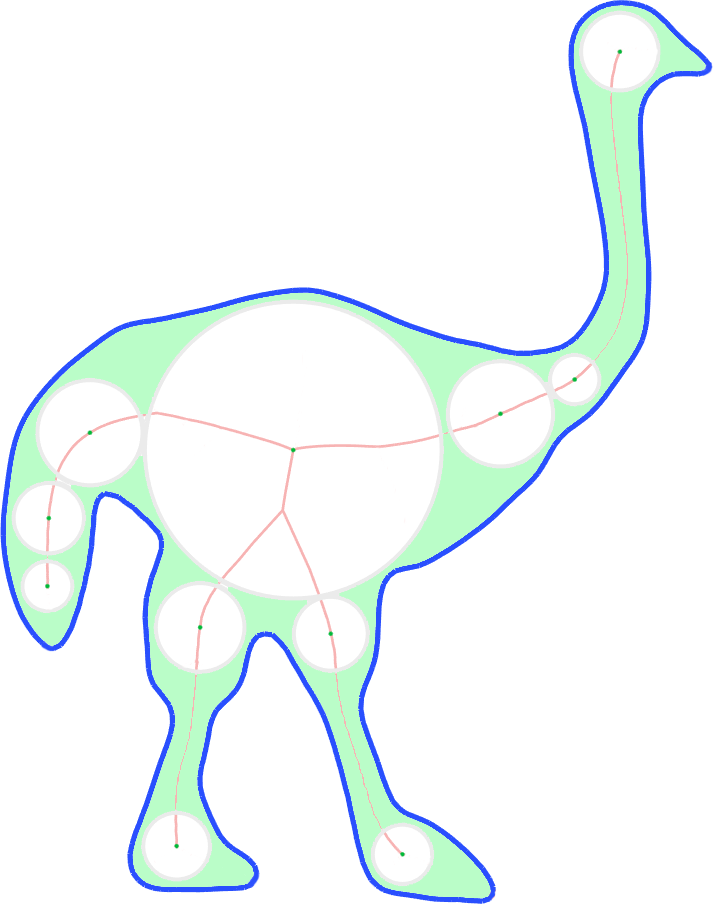}
             \makebox[0.03\linewidth][c]{}
    \includegraphics[width=0.36\linewidth]{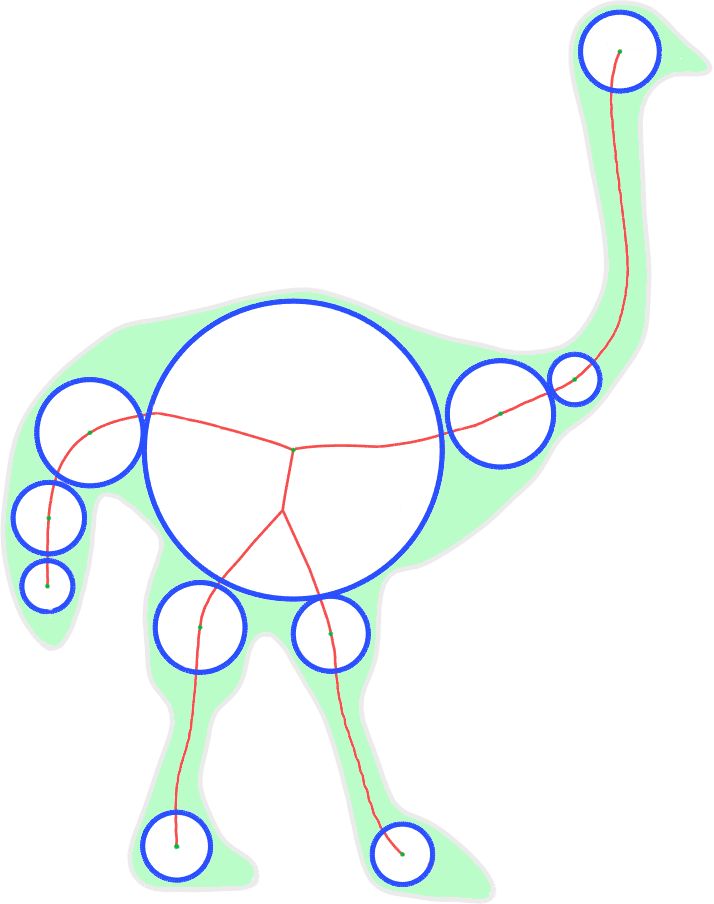}\\

    \vspace{0mm}
    \caption{Connecting selected poles based on Steiner tree. }
    \vspace{0mm}
    \label{fig:ball-stickmodel-Steiner}
\end{figure}

We have to connect the balls with wires to consolidate the structure of the toy.
Recall that each surface sampling point dominates a separate Voronoi cell restricted in the volume, and the union of the restricted Voronoi cells is said to be {\em restricted Voronoi diagram} (RVD).
The RVD naturally induces a graph $\mathcal{G}=(V,E)$ whose vertex set comprises the selected poles.
The key idea of bridging the selected poles is to find a Steiner tree belonging to $\mathcal{G}$ by taking each selected pole as a terminal.
For this purpose, we have to define a weighting scheme such that the tree with the highest centredness can be found. In this paper, we define the weight of a pair of adjacent MA points $p_i,p_j \in V$ as follows:
\begin{equation}\label{eq:weight}
  W_{ij}=\frac{\vert p_i-p_j \vert +\mu\vert r_i-r_j \vert}{\min(r_i,r_j)}.
\end{equation}
The parameter $\mu$ serves as a penalty to balance the two terms.
The term $\vert r_i-r_j \vert$ is to select a path along which the medial radius changes slowly,
the denominator $\min(r_i,r_j)$ is to keep the path away from the boundary surface,
and the term $\vert p_i-p_j \vert$ is to select an as-short-as-possible path.
After $P^+$ is fixed, we take the Steiner tree~\cite{garey1977rectilinear} of the graph $\mathcal{G}$ as the final connection structure between poles, where the selected poles $P^+\subset P \subset V$ are taken as terminals; See Figure~\ref{fig:ball-stickmodel-Steiner}.

\noindent
$\mathbf{Remark.}$~In fact, a Steiner tree is cycle free and thus can only be used to represent the skeleton of a genus-0 model. For a high-genus model, however, the skeleton is not a tree any more. We need to detect topology-aware loops first~\cite{erickson2005greedy}, and then with all the poles in $P^+$, together with the topologically characteristic loops, as terminals, we compute a connection structure that is structurally conformal to the boundary shape. Details are available in Appendix~B.

\subsection{Feature preserving}
\begin{figure}[ht]
    \centering
    \includegraphics[width=0.95\linewidth]{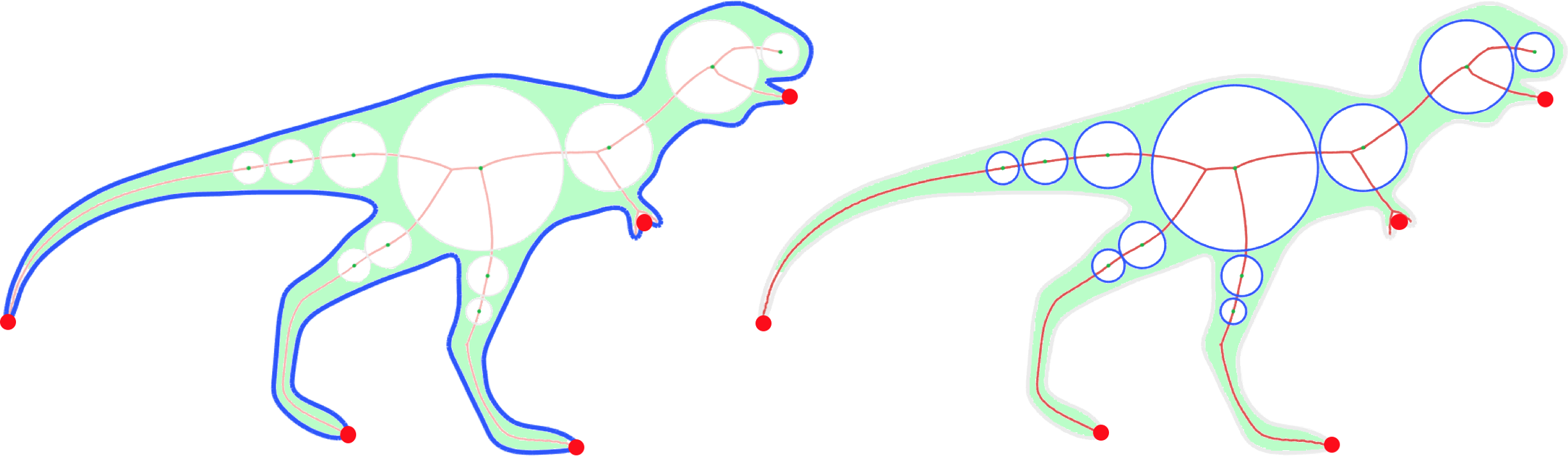}
    \vspace{1mm}
    \caption{Connecting selected poles and selected feature points.}
    \label{fig:ball-stickmodel-feature}
\end{figure}
Generally speaking, it is hard to use a sequence of balls to approximate the sharp features that are important hints to understand shapes.
To enable the ball-stick toy resemble the original input model as far as possible, we propose two strategies in the following.

\noindent \textbf{Separation gaps.}
In Eq.~(\ref{eq:selection2}), $d_{ij}\geq 0$ is a hard constraint to guarantee the separation between medial balls. In fact, we can further force the selected medial balls to be separated in a prescribed gap to facilitate the placement of wires.
\begin{equation}\label{eq:selection3}
 p^\text{new}=\mathop{\arg\max}_{p_j\in P^-}\{r_j\::\:d_{ij}\geq \epsilon, \forall p_i\in P^+\}.
\end{equation}
A side effect is to push the medial balls to around the sharp features. In spite of the difference between Eq.~(\ref{eq:selection2}) and Eq.~(\ref{eq:selection3}), the speedup technique mentioned in Section~\ref{subsec:speedup} still works with only a slight modification (we just change the medial radius of $p_j$ temporarily from $r_j$ to $r_j+\epsilon$ before updating $\mathbf{PD}(P^+\bigcup Q\bigcup p^\text{new})$).

\noindent \textbf{Feature points.} We further suggest adding significant feature points $p_i^f,i=1,2,\cdots,k,$ into the RVD-induced graph $\mathcal{G}$. Suppose that $p_i^f$ is located on the surface of the Voronoi cell of a certain sample point $s$.
We connect $p_i^f$ to any Voronoi vertex of $s$'s cell.
In this way, we augment $\mathcal{G}$ to $\mathcal{G}'$.
Similarly, we take the Steiner tree of $\mathcal{G}'$ to generate the final connection structure between selected poles and feature points; See Figure~\ref{fig:ball-stickmodel-feature} for illustration.

\subsection{Evaluation}\label{sec:evaluation}
\begin{figure}[ht]
    \centering
    \includegraphics[width=0.8\linewidth]{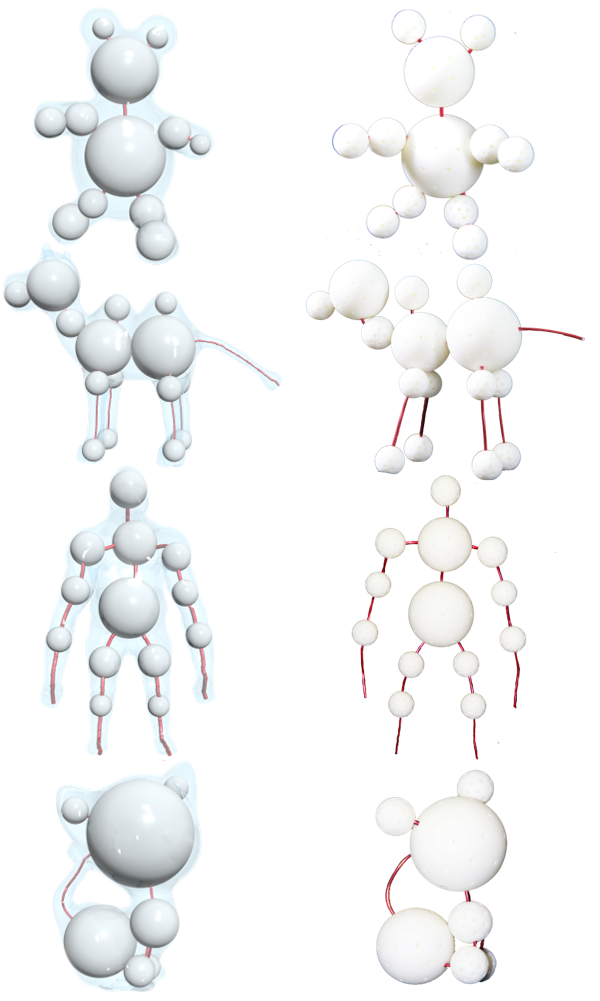}
    \vspace{0mm}
    \caption{More examples for building ball-stick toys. Teddy: 12 balls; Baby Camel: 15 balls; Gorilla: 13 balls; Kitten: 7 balls. Note that we customized four ball sizes, respectively 1.5cm, 2cm, 4.5cm, 5.5cm in diameter. \label{fig:ball-stick-examples}}
    \vspace{0mm}
\end{figure}
We implemented and experimented with our algorithm on a computer with a 64-bit version of macOS Mojave system, a 2.7 GHz Intel(R) Core(TM) i5 CPU and 16 GB memory. The coding language is C++. We employ the adaptive blue noise sampling approach~\cite{zhang2016capacity} in the sampling step. For simplicity, we remesh the Teddy model, the Baby-Camel model and the Gorilla model into 60K triangles and set the number of samples to be 4K. Besides, the program terminates when the radius of the top-priority ball popped from the priority queue is less than $5\%$ of the bounding box size. The parameter $\epsilon$ in Eq.~(\ref{eq:selection3}) is set to be $0.8\%$ of the bounding box size while the parameter $\mu$ in Eq.~(\ref{eq:weight}) is set to 3.

Our algorithm consists of four main operations. We use T1, T2, T3, T4 to respectively denote the four time costs:
\begin{itemize}
  \item[T1:] Compute Voronoi diagram w.r.t.~boundary samples;
  \item[T2:] Label inside and outside poles;
  \item[T3:] Extract RVD w.r.t.~boundary samples;
  \item[T4:] Connect poles in $P^+$.
\end{itemize}

The timing statistics (in seconds) for the models shown in Figure~\ref{fig:ball-stickmodel-Steiner},\ref{fig:ball-stickmodel-feature},\ref{fig:ball-stick-examples}  are available in Table~\ref{table:Ball-Stick}, where the \# Vertice denotes the number of boundary samples. In particular, we define 4 customized ball sizes, respectively 1.5cm, 2cm, 4.5cm, 5.5cm in diameter. Users can not only build their favorite ball-stick toys by reference to our generated results (see Figure~\ref{fig:ball-stick-examples}), but also transform it into various poses (see the teaser figure).

\begin{table}[ht]
\renewcommand\arraystretch{1.15}
\caption{Running time statistics for ball-stick toy design.~($\#$V denotes number of boundary samples)\label{table:Ball-Stick}}
\vspace{1mm}
\centering
\hspace*{-0.6em}
\scalebox{1.1}{
\begin{tabular}{|c|c|c|c|c|c|}
\hline
\multicolumn{2}{|c|}{Models} & \multicolumn{4}{c|}{Running Time (s)} \\ \hline
Name              & $\#$V     & T1      & T2      & T3      & T4      \\ \hline
Ostrich (2D)      & 500      & 0.004   & 0.007   & 0.002   & 0.139   \\ \hline
Dinosaur (2D)     & 500      & 0.005   & 0.007   & 0.003   & 0.127   \\ \hline
Teddy             & 4000     & 6.12    & 3.34    & 1.72    & 0.42    \\ \hline
Baby Camel        & 4000     & 6.19    & 2.86    & 2.99    & 0.525   \\ \hline
Kitten            & 4000     & 6.33    & 3.25    & 0.98    & 0.24    \\ \hline
Gorilla           & 4000     & 5.74    & 3.51    & 1.02    & 0.39    \\ \hline
\end{tabular}}
\end{table}

\begin{figure}[!ht]
    \centering
    \includegraphics[width=0.86\linewidth]{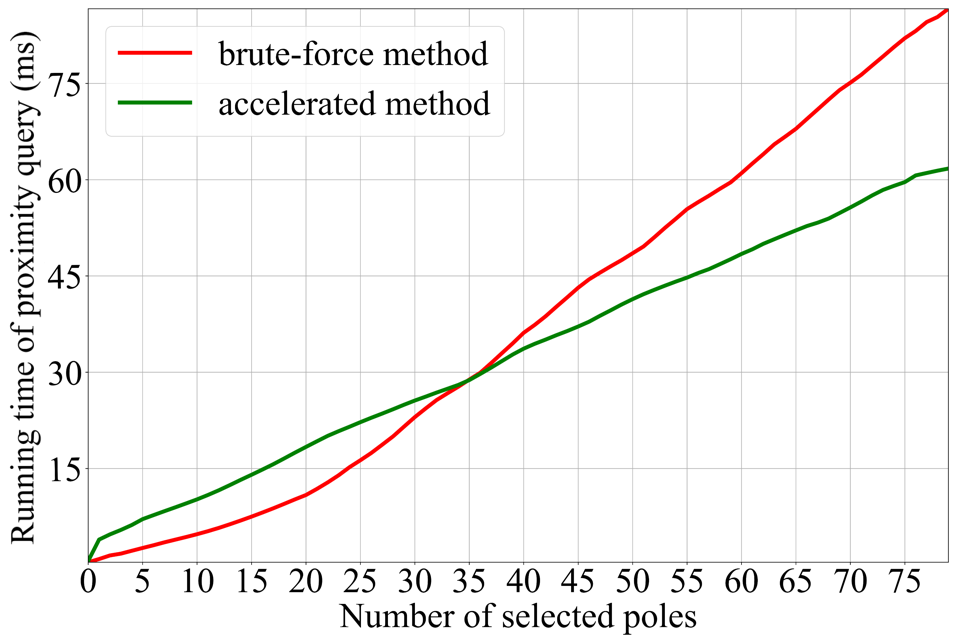}\\
    \caption{Total running time v.s. the number of selected poles.\label{fig:time_comp}}
\end{figure}

In the following, we use Figure~\ref{fig:time_comp} to show the performance contrast between the brute-force ball penetration detection strategy (Algorithm~1) and the accelerated strategy (Section~~\ref{subsec:speedup}). The test is conducted on the Kitten model with 4K points being sampled.

Without doubt, when just a few poles are selected, the brute-force ball penetration detection strategy may be faster. But with the increasing of the number of selected poles, the accelerated strategy outperforms the nai\"ve one. As shown in Figure~\ref{fig:time_comp}, when the number of selected poles exceeds $35$, our accelerated method exhibits a conspicuous advantage.

\begin{figure*}
\centering
 \includegraphics[width=0.19\linewidth]{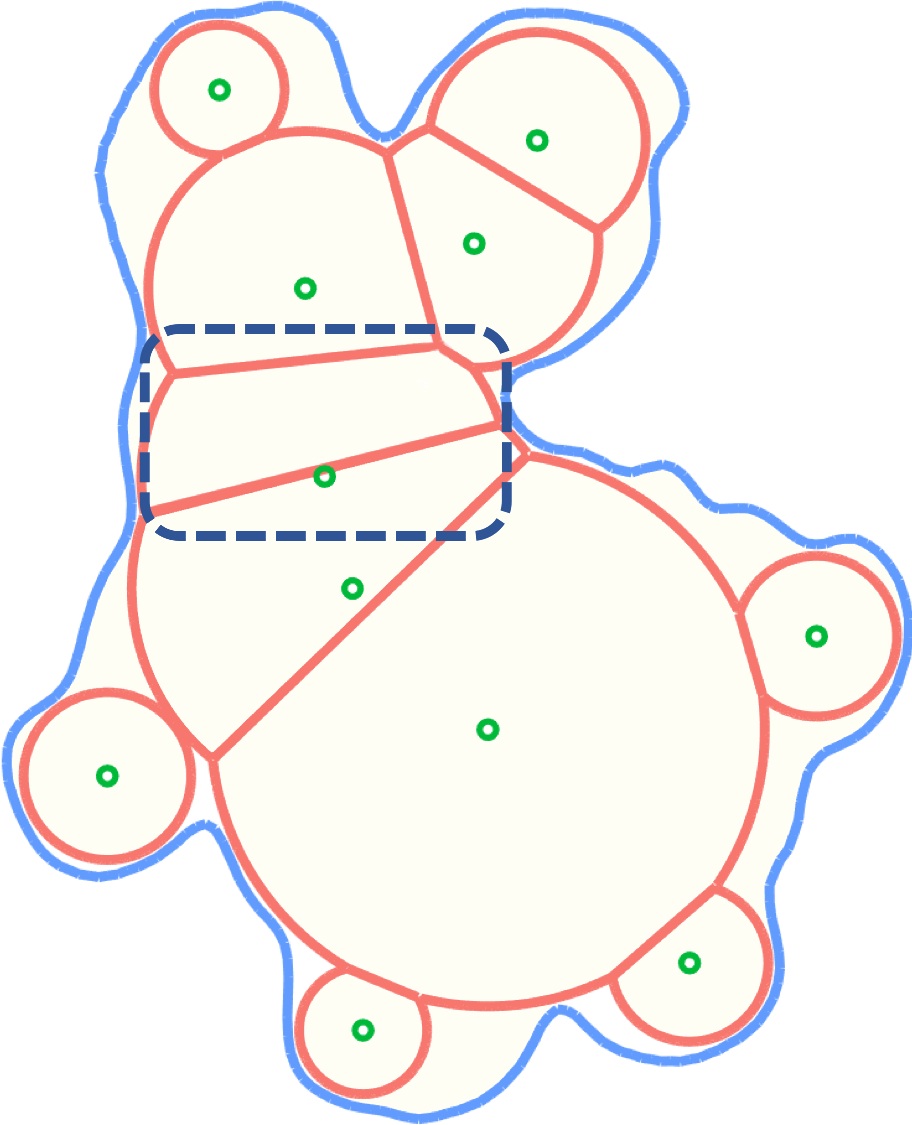}
 \includegraphics[width=0.19\linewidth]{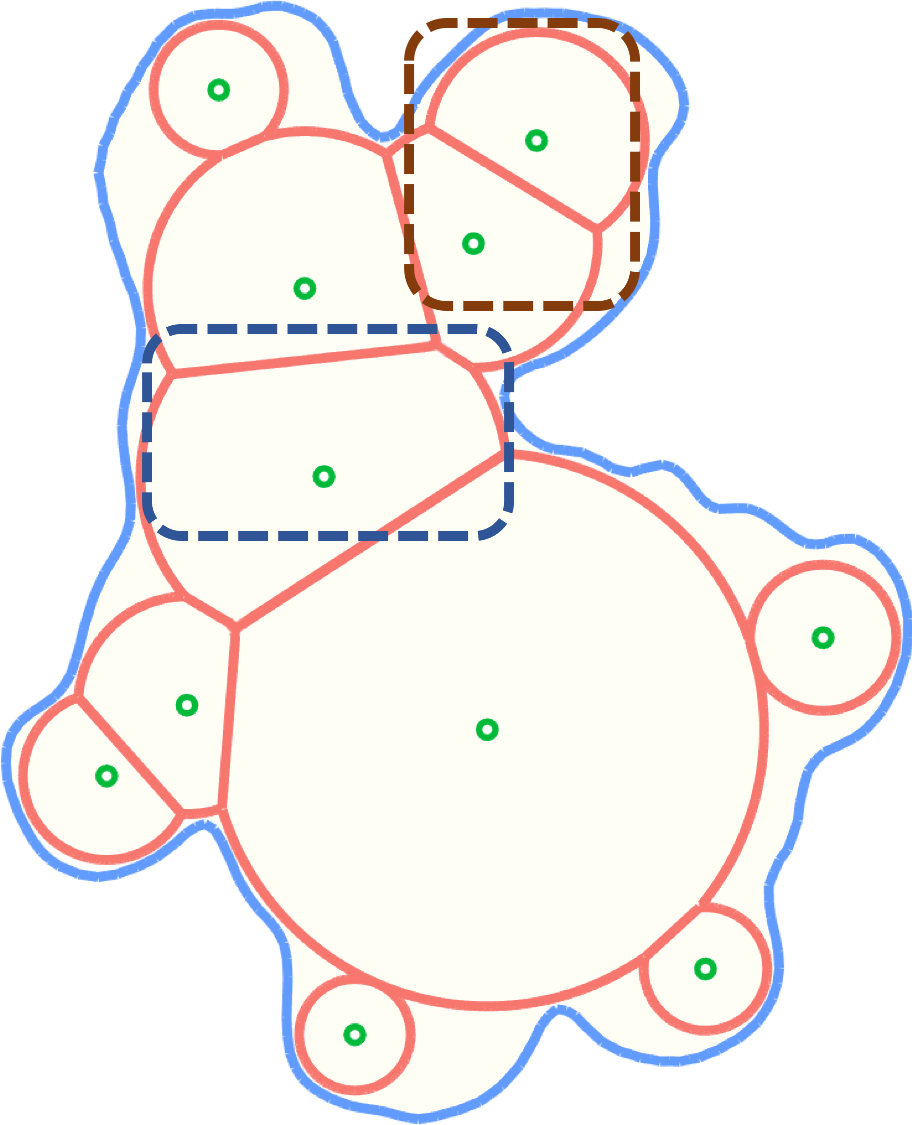}
 \includegraphics[width=0.19\linewidth]{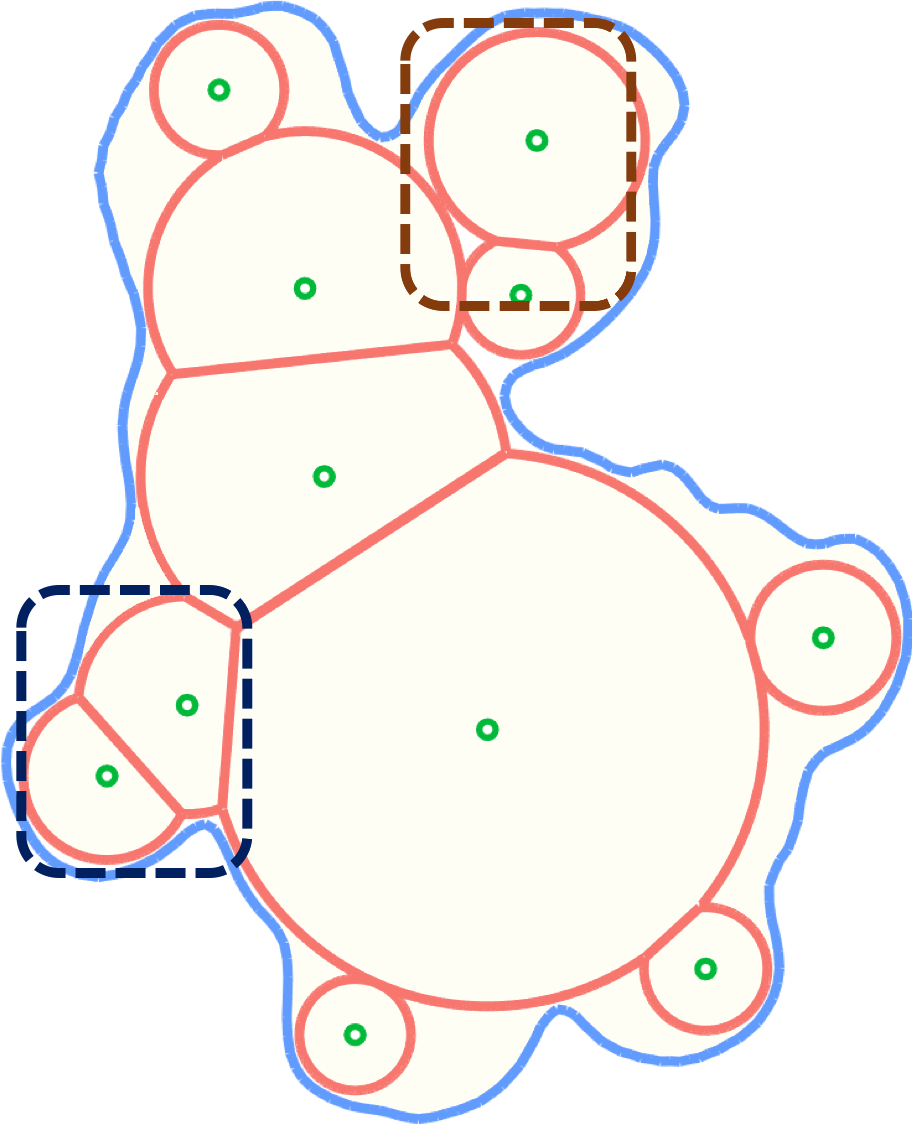}
 \includegraphics[width=0.19\linewidth]{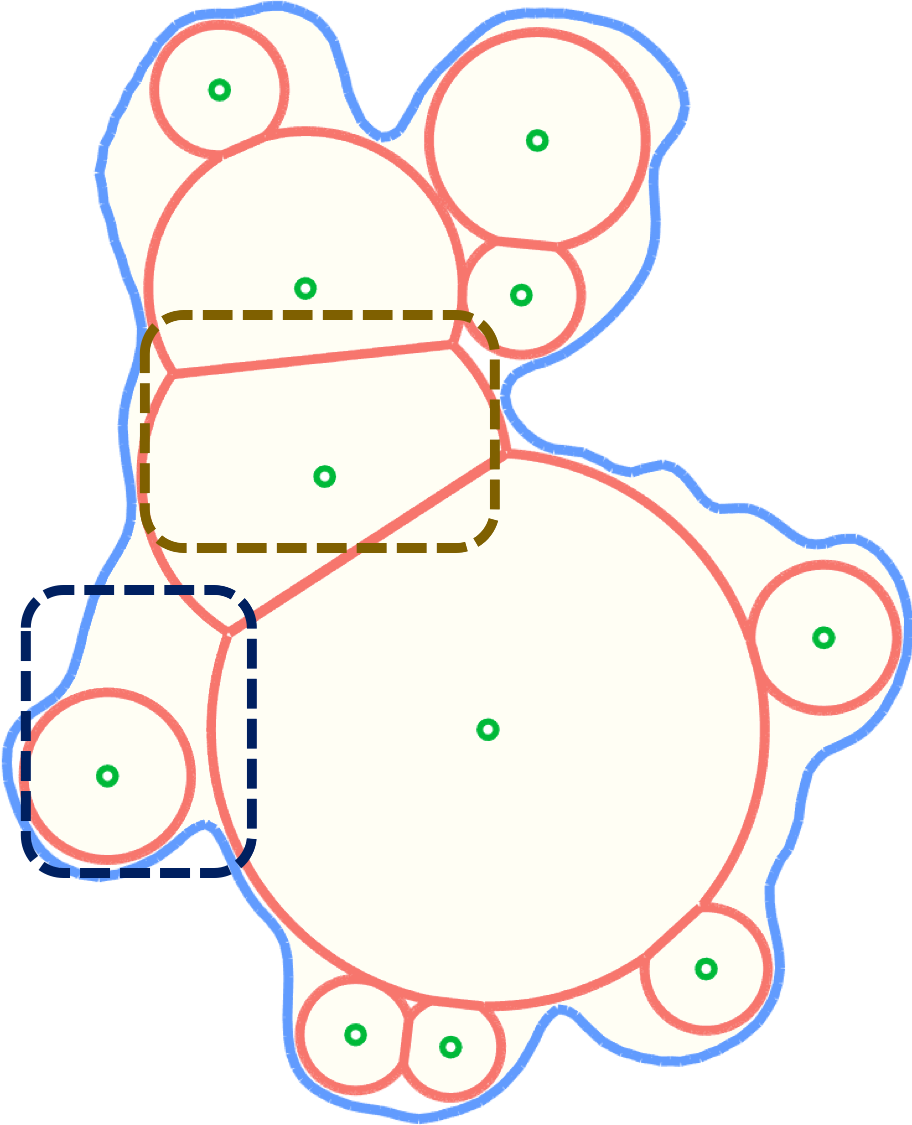}
 \includegraphics[width=0.19\linewidth]{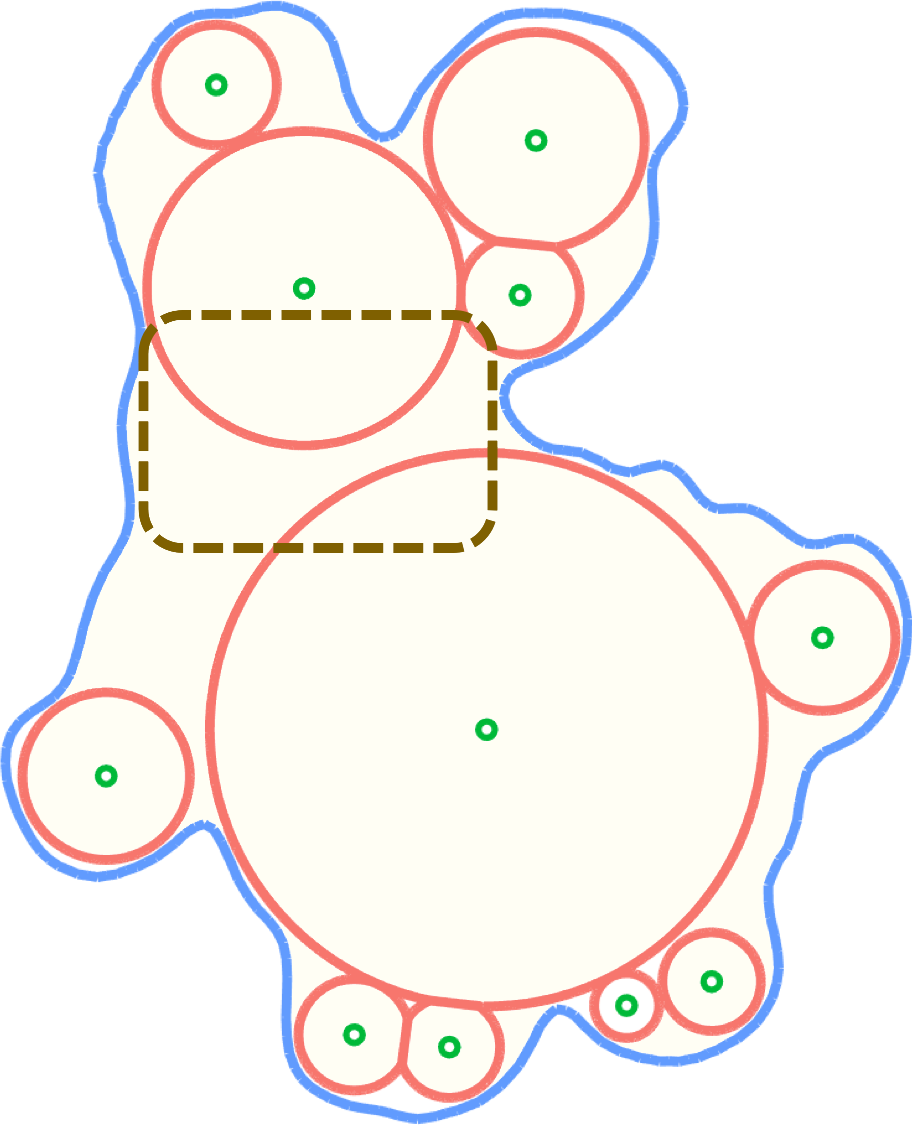}
   {
   \leftline{\makebox[1.0\linewidth][l]{\quad \quad \quad \; \,
   (a)~$\lambda = 0.5$ \quad \quad \quad \quad \quad \,
   (b)~$\lambda = 0.6$ \quad \quad \quad \quad \quad \; \,
   (c)~$\lambda = 0.7$ \quad \quad \quad \quad \quad \,
   (d)~$\lambda = 0.8$ \quad \quad \quad \quad \quad \quad
   (e)~$\lambda  = 0.95$}
  }}
  \vspace{-5mm}
 \caption{The influence of the parameter $\lambda$ - a larger value of $\lambda$ tends to suppress penetration between medial balls. The differences between two successive results are highlighted. We choose $\lambda$ to be 0.8 in our experiments.\label{fig:lambda}}
\end{figure*}

\section{Porous Structure Generation}\label{sec:porous}
\begin{wrapfigure}{r}{1.8cm}
  \vspace*{-4mm}
  \hspace*{-3mm}
  \centerline{
  \includegraphics[width=2cm]{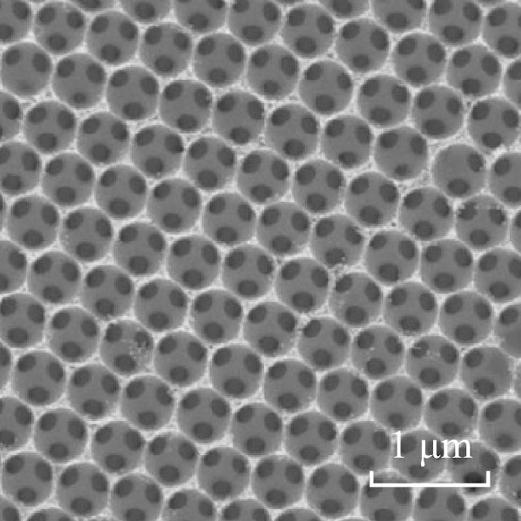}
  }
  \vspace*{-4mm}
\end{wrapfigure}
In the following, we shall discuss the other application, i.e., how to generate porous structures.
Porous structures are ubiquitous in nature due to their nice physical properties.
Porous structure design is a fascinating yet challenging research topic in recent years, especially with the innovation of the additive manufacturing technology.
As the inset figure shows, the spherical inverse opal is the most common processed and studied self-assembly structure~\cite{ma2014si}. Compared with other pore structures such as ellipsoids, plates and rods, the spherical inverse opal has better structural mechanical behavior. This motivates us to generate shape-aware spherical porous structures.

\subsection{Requirements}

We intend to select a set of medial balls to generate porous structures. Before we figure out detailed selection strategies, we made the following two experiments to make clear (1)~whether spherical pores are better than pores of other types, and (2)~to what degree the penetration between pores reduces the entire mechanical rigidity.
\begin{figure}[!h]
    \centering
       \includegraphics[width=0.75\linewidth]{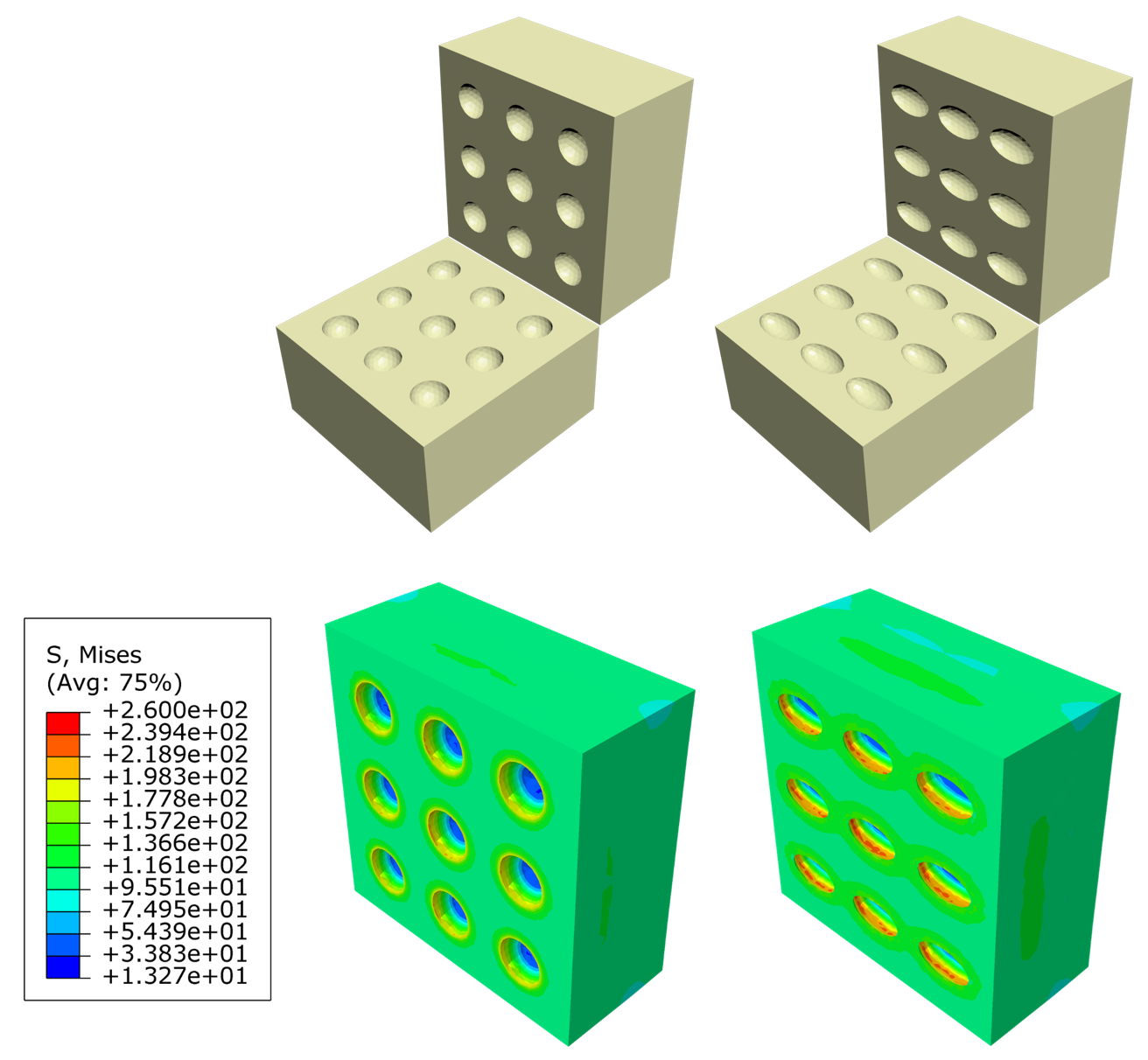}
       \vspace{-1mm}
    \makebox[0.5\linewidth][r]{3$\times$3$\times$3}\makebox[0.5\linewidth][l]{ \,  \quad \quad \quad \quad 3$\times$3$\times$3}
    \makebox[0.5\linewidth][r]{spherical pores }\makebox[0.35\linewidth][c]{ellipsoidal pores}
    \vspace{-1mm}
    \caption{The stress nephograms in ABAQUS show that under condition that equal-volume material is removed from a cube, spherical pores own better mechanical rigidity than ellipsoidal pores since the maximum stress for the spherical-pore structure is 214MPa while the maximum stress for the ellipsoidal-pore structure is 253.1MPa. }
    \label{fig:convexity}
\end{figure}

\noindent
$\mathbf{Convexity/compactness.}$~As Figure~\ref{fig:convexity} shows, we generate a spherical-pore structure and an ellipsoidal-pore structure respectively for the input cube model, and the pores in both configurations are of the same volume size. From the stress nephograms in ABAQUS, we can clearly see that spherical pores are able to better preserve the original mechanical rigidity, which is in line with our initial guess. Therefore we intend to select some representative medial-axis balls as basic primitives to compose the final pore structure.

\begin{figure}[h]
    \centering
    \hspace*{-2mm}
       \includegraphics[width=1.0\linewidth]{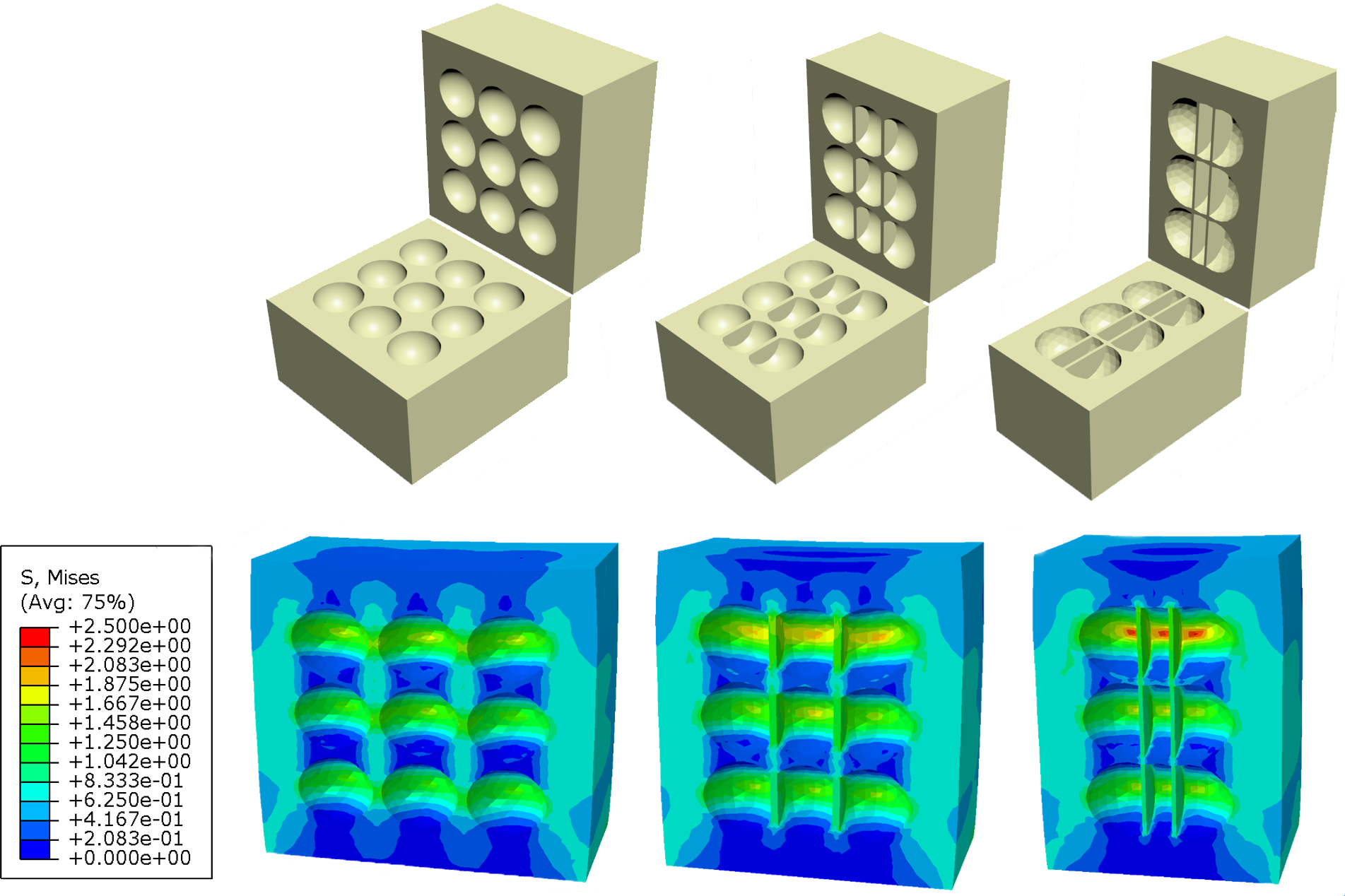}\\
      \vspace{-1mm}
       {\flushleft
    \makebox[1.0\linewidth][l]{ \, \quad \quad \quad \quad  \quad \quad \quad \quad (a)
    \quad \quad \quad \quad \quad \, (b)
    \quad \quad \quad \quad \quad \, (c)}}
    \vspace{-2.5mm}
    \caption{By enforcing top-down pressure, we find that the stress distribution for the model in (b), with moderate penetration, is rather uniform while the weight saving rate amounts to $20\%$, which is higher than (a) and (c). Note that we tune the wall thickness of (a-c) to the same size to facilitate observing the effect of pore penetration.  }
    \label{fig:Penetration}
    \vspace{0mm}
\end{figure}

\noindent
$\mathbf{Penetration~between~medial~balls.}$~Different from the ball-stick toy design, penetration between two medial balls should be allowed as long as the convexity/compactness of each pore is not seriously destroyed. In Figure~\ref{fig:Penetration}, we show three configurations of pores, i.e. no penetration, moderate penetration and serious penetration. The stress nephograms show that the pore configuration shown in Figure~\ref{fig:Penetration}(b) saves more material but can well preserve the entire mechanical rigidity. Therefore, we allow moderate intersection between medial balls, rather than enforce a hard constraint of being penetration-free.
\subsection{Selection strategy}
Based on the above experiments, we consider the selection strategy from two aspects.
On one hand, we intend to encourage those large medial balls to be selected into $P^+$, and on the other hand, we have to enforce a penalty on serious penetration, i.e.,
\begin{equation}\label{eq:selection-porus}
p^\text{new}=\mathop{\arg\max}_{p_j\in P^-}\{r_j+\lambda\times\min\big(0,\min_{p_i\in P^+}d_{ij}\big)\},
\end{equation}
where $0<\lambda < 1$ is used to balance the two considerations. Similarly, we stop the selection process if the medial radius of $p^\text{new}$ is sufficiently small.

\noindent
$\mathbf{Remark.}$~If $\lambda$ is small, then there may be many thin pores generated due to lack of penalty on penetration; See Figure~\ref{fig:lambda}.

If $\lambda=1$, the to-be-selected pole must be located outside the existing balls.
W.l.o.g, we take each surface sample point as a pole whose medial radius is 0.
Obviously, all the sample points are located outside the existing medial balls.
Therefore, it is easy to infer that $$r_j+\min\big(0,\min_{p_i\in P^+}d_{ij}\big)\geq 0$$
for any boundary sample point.
This implies that under the selection criterion $$p^\text{new}=\mathop{\arg\max}_{p_j\in P^-}\{r_j+\min\big(0,\min_{p_i\in P^+}d_{ij}\big)\},$$ the to-be-selected pole $p_j$ must meet $r_j+\min\big(0,\min_{p_i\in P^+}d_{ij}\big)\geq 0 $ or at least $|p_j-p_i|-r_i \geq 0.$ So $p_j$ must be located outside the existing balls. In our experiments, we set $\lambda=0.8$ without any specification.

\subsection{Implementation}
In the following, we give a couple of key technical details of algorithm implementation.

\noindent
$\mathbf{Dynamic~power~diagram.}$~
Throughout the algorithm, we need to maintain two data structures. One is the priority queue where a pole $p_j\in P^-$ has a priority of $r_j+\lambda\times\min\big(0,\min_{p_i\in P^+}d_{ij}\big)$, and the other is a dynamic power diagram w.r.t.~$P^-$.

Imagine that $p^\text{new}\in P^-$ is the newly selected pole that is ready to be added to $P^+$. According to Eq.~(\ref{eq:selection-porus}), the selection of $p^\text{new}$ possibly changes the priority of those poles in $P^-$ that are very close to  $p^\text{new}$ (i.e., penetration occurs between $p^\text{new}$'s medial ball and $p_j$'s medial ball). We need to quickly identify these poles and update their priorities.

Once again, we employ Theorem~\ref{thm:interference} to achieve this purpose. Let $\mathbf{PD}(Q\bigcup P^-)$ be the power diagram at this moment, where $Q$ is the outside pole set. Let $\widehat{P^-}$ be a copy of $P^-$. We repeatedly label and then remove a pole $p_j\in \widehat{P^-}$ as long as $p_j$'s cell is neighboring to $p^\text{new}$'s cell in the up-to-date power diagram. We stop the process until $p^\text{new}$'s cell becomes isolated from those remaining poles in $\widehat{P^-}$ (i.e. the boundary of $p^\text{new}$'s cell is given by $p^\text{new}$ and the outside poles).
So far, we have identified a set of poles in $P^-$ whose medial balls possibly intersect with $p^\text{new}$'s medial ball. We then update the priority of these identified poles.
Finally we compute $\mathbf{PD}( Q\bigcup P^-\setminus p^\text{new})$ based on a light-weight update of $\mathbf{PD}(Q\bigcup P^-)$ and continue the next iteration.

\begin{figure*}
\centering
 \includegraphics[width=1.0\linewidth]{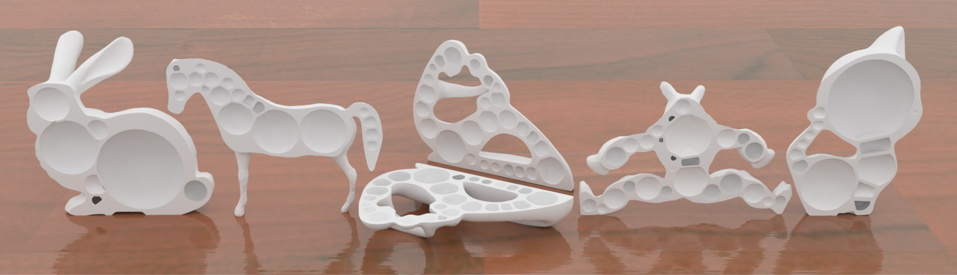}
  \vspace{-3mm}
 \caption{Section views of spherical-pore structures. Bunny: 17 pores; Horse: 30 pores; Fertility: 66 pores; Armadillo: 42 pores; Kitten: 23 pores.\label{fig:porous-examples}}
 \vspace{-1mm}
\end{figure*}

\noindent
$\mathbf{Thickness~control.}$~
We need to keep the pores away from the boundary surface in a proper gap. At the same time, a planar separator between neighboring pores is used to preserve rigidity. We introduce a parameter $\tau$ to achieve this purpose.

First, we filter out those inside poles whose medial radius is less than $\tau$.
We then reduce the medial radii of other inside poles by $\tau$.
In fact, the pore belonging to $p_i\in P^+$ can be obtained by subtracting the medial ball $B(p_i,r_i-\tau)$ by a sequence of equal-power planes, each of which is given by $p_i$ and a neighboring site $p_j\in P^+$ (here ``neighboring'' means that penetration occurs between $p_i$'s medial ball and $p_j$'s medial ball). In order to guarantee the separator being of a $\tau$-thickness, we move each equal-power plane in a distance of $\frac{\tau}{2}$ toward both sides upon penetration between medial balls occurs.

\subsection{Evaluation}
\noindent
$\mathbf{Performance.}$~
Our algorithm involves three main operations and we use T1, T2, T3 to respectively represent their time costs.
 T1: the time cost for computing Voronoi Diagram w.r.t.~boundary samples; T2: the time cost for labeling inside and outside poles;
  T3: the time cost for computing the power diagram w.r.t.~$P^+$. We use totally five models for test; See section views of spherical-pore structures in Figure~\ref{fig:porous-examples}.
  From the timing statistics in Table~\ref{table:porous-performance} we can see that our algorithm runs in only tens of seconds for 4K samples.
Note that generally T3 is related to the number of poles in~$P^+$, which accounts for why T3 is larger for the Fertility model than other models.

\begin{table}[h]
\renewcommand\arraystretch{1.15}
\caption{Running time statistics for porous structure generation.\label{table:porous-performance}~($\#$V denotes number of boundary samples)}
\vspace{1mm}
\centering
\hspace*{-0.6em}
\scalebox{1.1}{
\begin{tabular}{|c|c|c|c|c|c|}
\hline
\multicolumn{2}{|c|}{Models} & \multicolumn{3}{c|}{Running Time (s)} & \multirow{2}{*}{\begin{tabular}[c]{@{}c@{}}Weight Sav-\\ ing Ratio\end{tabular}} \\ \cline{1-5}
Name         & $\#$V  & T1         & T2         & T3          &                                                                                \\ \hline
Kitten       & 4000          & 5.91       & 2.11       & 6.62        & 62.85\%                                                         \\ \hline
Fertility    & 4000          & 6.09       & 3.23       & 20.67       & 57.71\%                                                                        \\ \hline
Horse        & 4000          & 6.29       & 2.82       & 7.79        & 51.86\%                                                                        \\ \hline
Bunny        & 4000          & 5.44       & 3.09       & 4.65        & 54.82\%                                                                     \\ \hline
Armodillo    & 4000          & 5.95       & 2.65       & 12.43       & 46.95\%                                                                        \\ \hline
\end{tabular}
}
\end{table}

\begin{table}[h]
\renewcommand\arraystretch{1.12}
\vspace{-2mm}
\caption{Pressure test statistics.\label{fig:ratio-cup}}
\vspace{1mm}
\centering
\hspace*{-0.6em}
\scalebox{1.08}{
\begin{tabular}{|c|c|c|}
\hline
Item                                                                                   &  \begin{tabular}[c]{@{}c@{}}Spherical-pore\end{tabular} & Honeycomb \\ \hline
Solid Vol (cm$^3$)                                                                        & 540.88         & 540.88    \\ \hline
Actual Material Vol (cm$^3$)                                                              & 213.08         & 226.79    \\ \hline
Max Load~(kN)                                                                     & 0.64           & 0.49      \\ \hline
\begin{tabular}[c]{@{}c@{}}Max Load / Material\\ Vol Ratio (N/cm$^3$)\end{tabular} & 3.00           & 2.16         \\ \hline
\end{tabular}
}
\end{table}

\begin{figure}[!h]
    \centering
    \includegraphics[width=0.86\linewidth]{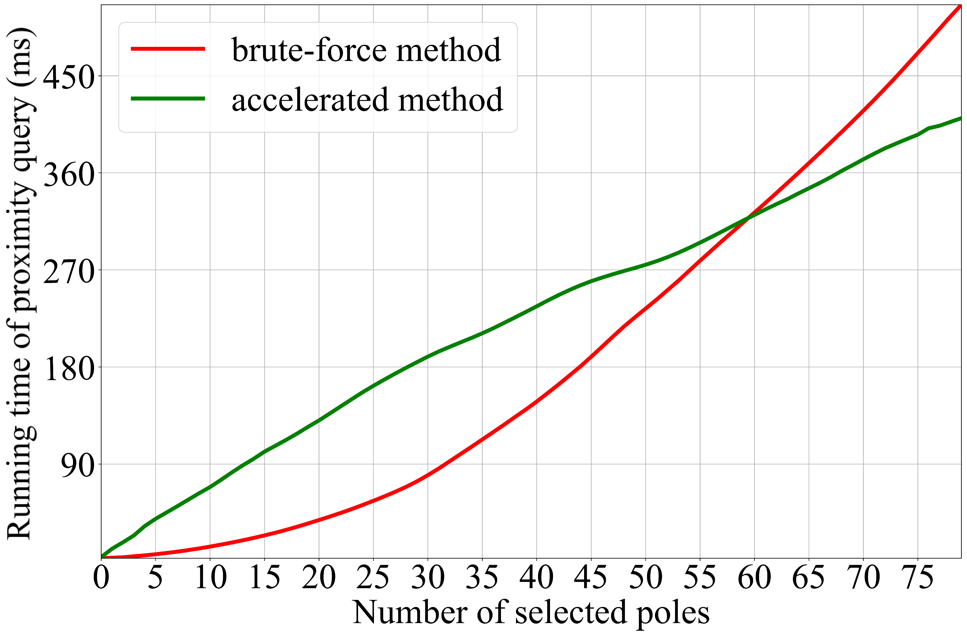}\\
    \vspace{0mm}
    \caption{Total running time v.s. the number of selected poles.\label{fig:time_comp_porous}}
    \vspace{0mm}
\end{figure}

In the following, we discuss how effective our priority update strategy is.
Let $p$ be a newly selected pole.
In fact, a na\"{i}ve strategy of updating the priorities of the remaining poles is to detect the penetration between $p$ and all the poles in $P^-$. By contrast, our accelerated strategy is to update the priorities of just a few surrounding inside poles, instead of detecting every unselected poles in a brute-force manner.
We conducted a test on the Kitten model with 4K samples. The performance statistics are given in Figure~\ref{fig:time_comp_porous}.

\begin{figure}[!h]
\centering
 \includegraphics[width=1\linewidth]{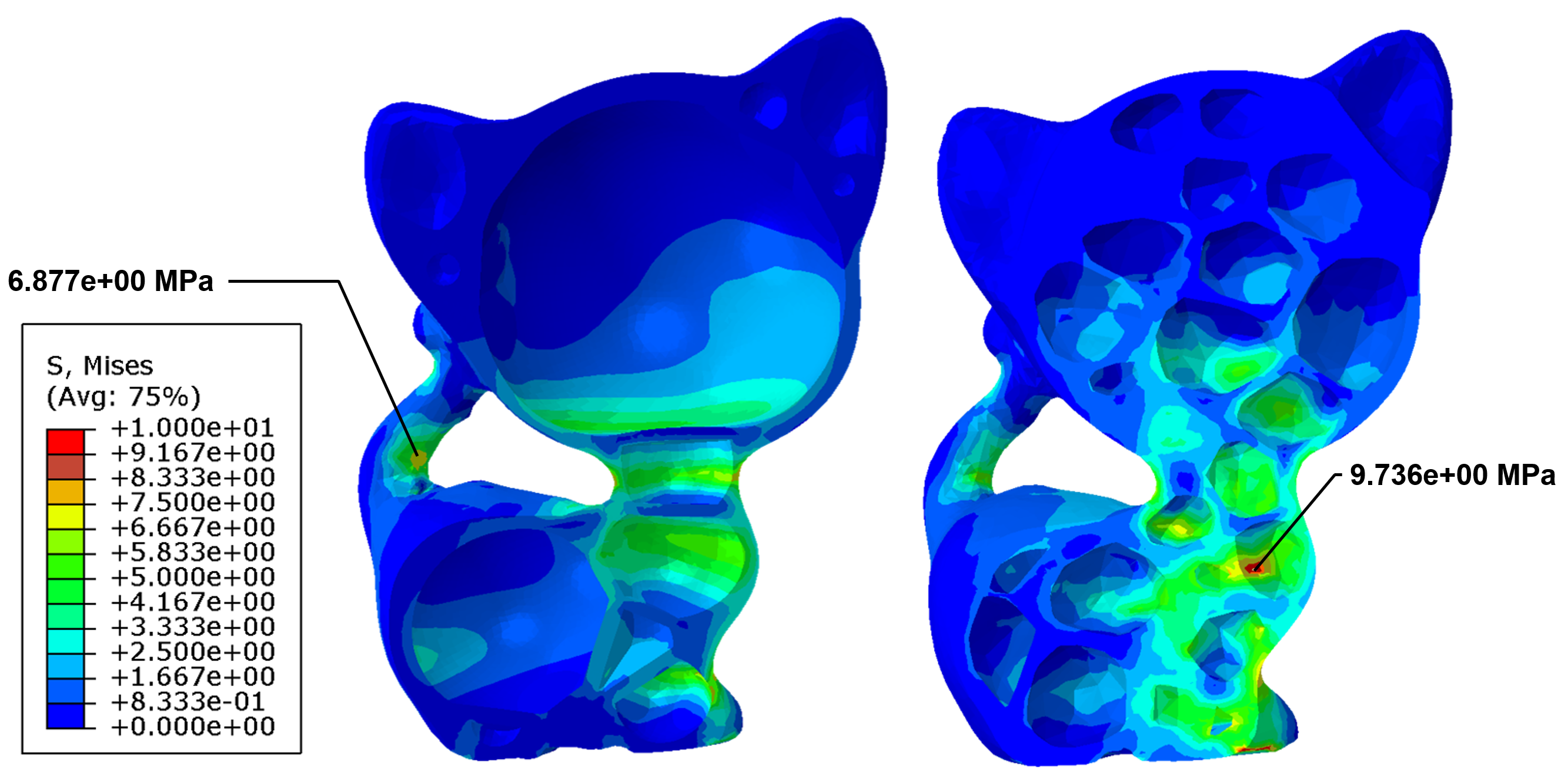}\\
 \vspace{0mm}
 \makebox[0.5\linewidth][r]{(a)~Spherical pores}\makebox[0.5\linewidth][c]{(b)~Honeycomb structure}
  \vspace{-2mm}
 \caption{Finite element analysis for the spherical-pore Kitten model and the honeycomb-structure Kitten model~\cite{lu2014build}.
  (a)~By enforcing a gravity load, the FEA result shows that the spherical-pore Kitten model (this paper) owns a more uniform distribution of stress with a weight saving rate of $63\%$. (b)~The honeycomb-structure Kitten model has a weight saving rate of $58\%$ but its press distribution is non-uniform.
  \label{fig:kitten-fea}}
  \vspace{-1mm}
\end{figure}

\noindent
$\mathbf{Finite~element~analysis.}$~
In order to compare with the honeycomb-structure hollowing method~\cite{lu2014build}, we generate two versions of the Kitten model to perform finite element analysis by enforcing a gravity load. The paramter configurations are as follows: Young's modulus, Poisson's ratio, mass density are respectively set to 2100, 0.31, 3$\times 10^{-9}$. The resulting stress nephogram in ABAQUS shows that our spherical-pore Kitten model, in spite of larger weight saving ratio, owns a more uniform distribution of stress. See Figure~\ref{fig:kitten-fea}.
Note that red means that the stress is high, while blue stands for low stress.
It can be seen that the maximum stress of the honeycomb structure occurs around the neck, amounting to 9.736Mpa,
which is due to two facts: (a)~the honeycomb cells' surfaces have varying curvatures, and are much different from curvature-uniform spherical pores, and (b)~the neck is the weakest part and thus should not be reduced too much in weight.

\begin{figure}[ht]
\centering
 \includegraphics[width=0.99\linewidth]{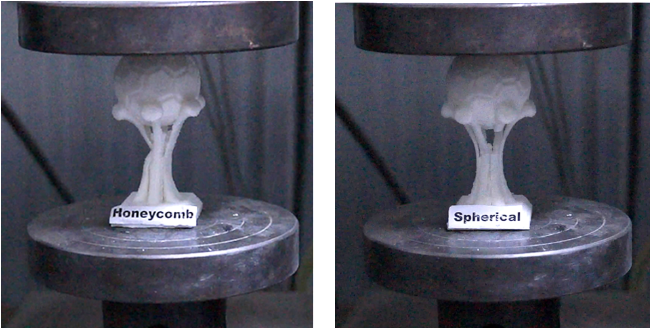}\\
 \makebox[0.5\linewidth][c]{(a)~Honeycomb structure}\makebox[0.5\linewidth][c]{(b)~Spherical pores}
 \vspace{-3mm}
 \caption{ The values of compaction force at the breaking point are 0.49kN (a) and 0.64kN (b) respectively, which shows that our resulting model has a better mechanical rigidity. \label{fig:breaking}}
 \vspace{0mm}
\end{figure}

\noindent
$\mathbf{Physical~tests.}$~
We used a hydraulic testing machine to report the maximum pressure the model can withstand.
We downloaded the honeycomb-structured Trophy model from the project page of~\cite{lu2014build}. Their theoretical weight saving rate is $58.1\%$. For the comparison purpose, we generated a spherical-pore structure with a theoretical weight saving rate of $60.6\%$ by controlling the number of pores. With PLA+ as the 3D printing material (tensile strength:~60Mpa, bending strength:~87Mpa, bending modulus:~3642Pa), we used an FDM prototyping machine to print the two kinds of porous structures for pressure test.
Detailed pressure statistics are available in Table~\ref{fig:ratio-cup}.
Figure~\ref{fig:breaking} captured the moments when the model breaks, and the values of compaction force at the breaking point are 0.64kN (ours) and 0.49kN (honeycomb-structured) respectively. Note that the significant difference in structure is that the honeycomb-structured model was hollowed even around the thin-neck parts, but our algorithm didn't produce any spherical pores around the neck.
Therefore we believe that our proposed algorithm has a big potential in the light-weight product design.

\section{Conclusion}
In this paper, we studied the medial ball based shape abstraction problem, i.e., how to use as few medial balls as possible to approximate the original enclosed volume while imposing given geometric constraints. Our technical contribution is to
propose a top-down selection strategy to encourage large medial balls while taking into account specific requirements.
We develop an effective speedup technique based on a provable observation that the intersection of medial balls implies the adjacency of power cells (in the sense of the power crust).

We further explore the uses of our algorithm in combination with two closely related applications. In the ball-stick design application, our algorithm is able to help non-professional users to quickly build a shape with only balls and wires, where any penetration between two medial balls must be suppressed. In the porous structure generation application, however, moderate penetration between two adjacent spherical pores are allowed to maximize the ratio of material saving while preserving structural rigidity. We use both FEA and real physical tests to validate the effectiveness.

\ifCLASSOPTIONcompsoc
  \section*{Acknowledgments}
\else
  \section*{Acknowledgment}
\fi

The authors would like to thank the anonymous reviewers for their valuable comments and suggestions. This work is supported by National Natural Science Foundation of China (61772016, 61772318) and Research Grant Council of Hong Kong (GRF 17263316).

\ifCLASSOPTIONcaptionsoff
  \newpage
\fi

\bibliographystyle{IEEEtran}
\bibliography{ourbibfile}
\vspace{-8mm}
\begin{IEEEbiography}[{\includegraphics[width=1in,height=1.25in,clip,keepaspectratio]{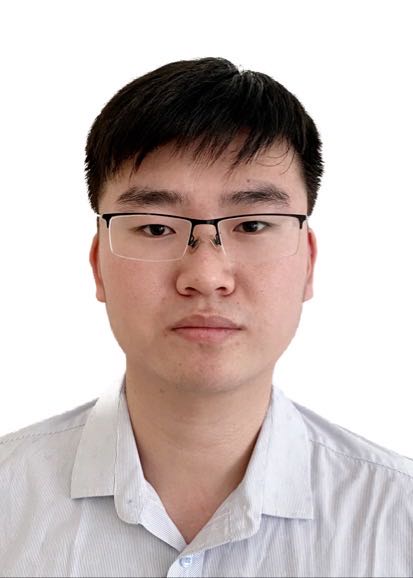}}]{Zhiyang Dou} is a 4th-year student in the School of Computer Science and Technology, Shandong University, China. His current research interests include geometric processing and computational geometry.
\end{IEEEbiography}

\begin{IEEEbiography}[{\includegraphics[width=1in,height=1.25in,clip,keepaspectratio]{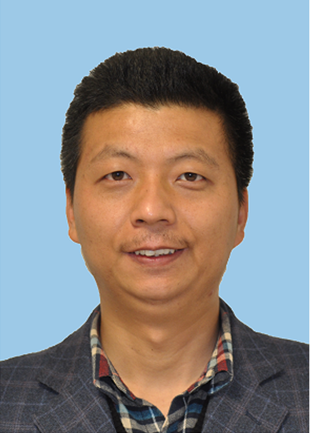}}]{Authors/Shiqing Xin} is an associate professor within the school of computer science at Shandong University. He got Ph.D. at Zhejiang University (China) in 2009. After that,he worked as a research fellow at Nangyang Technological University (Singapore) for three years. His research interests include various geometry processing algorithms, especially geodesic computation approaches and Voronoi/power tessellation methods. During the past ten years, he published over 60 papers on famous journals/conferences, including IEEE TVCG, ACM TOG, etc. He got three Best Paper awards and many other academic awards.
\end{IEEEbiography}

\begin{IEEEbiography}[{\includegraphics[width=1in,height=1.25in,clip,keepaspectratio]{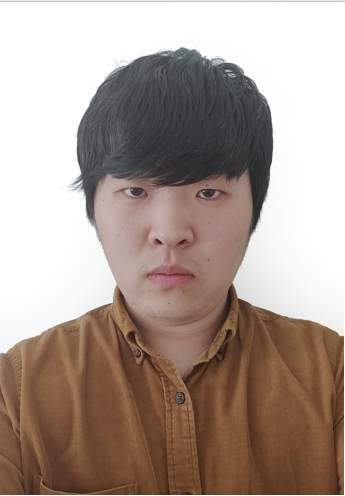}}]{Authors/Rui Xu} is a 3rd-year student in the School of Computer Science and Technology, Shandong University, China. His research interests include geometric processing and computational geometry.
\end{IEEEbiography}
\vfill

\begin{IEEEbiography}[{\includegraphics[width=1in,height=1.25in,clip,keepaspectratio]{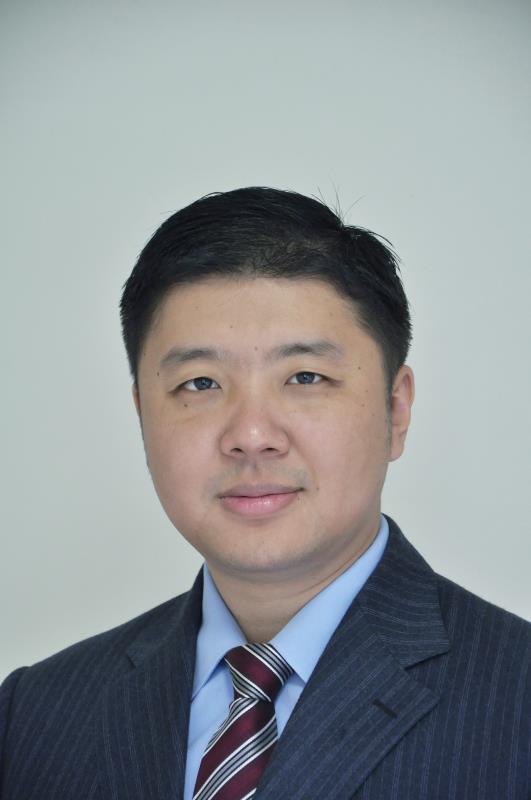}}]{Jian Xu} received bachelor's, master's and Ph.D. degrees in Ocean University of China, Tu-Braunschweig (Germany) and Catholic University of Leuven (Belgium) respectively. He is now a senior researcher of Ningbo Institute of Materials Technology and Engineering (Chinese Academy of Sciences), and professor of Dalian University of technology. Prof. Xu has been focusing on composite equipment development, composite process simulation, full-scale composite simulation and composite process development. Totally 35 papers were published.
\end{IEEEbiography}

\begin{IEEEbiography}[{\includegraphics[width=1in,height=1.25in,clip,keepaspectratio]{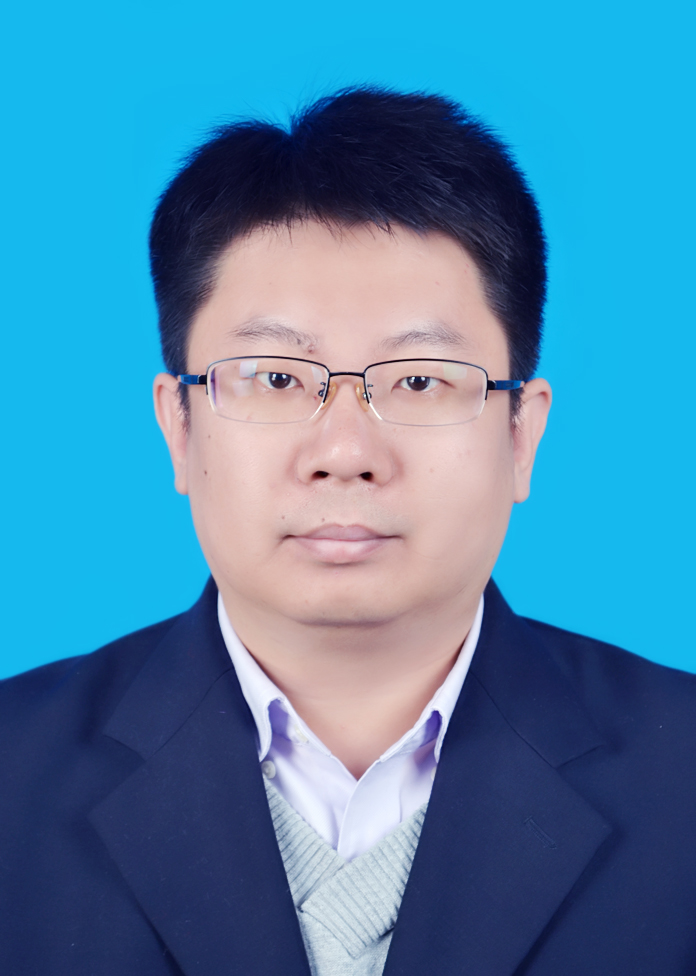}}]{Yuanfeng Zhou} received the master's and Ph.D. degrees from the School of Computer Science and Technology, Shandong University, Jinan, China, in 2005 and 2009, respectively. He held a post-doctoral position with the Graphics Group, Department of Computer Science, The University of Hong Kong, Hong Kong, from 2009 to 2011. He is currently a professor with the School of Software, Shandong University, where he is also a member of the GDIV Laboratory. His current research interests include geometric modeling, information visualization, and image processing.
\end{IEEEbiography}

\begin{IEEEbiography}[{\includegraphics[width=1in,height=1.25in,clip,keepaspectratio]{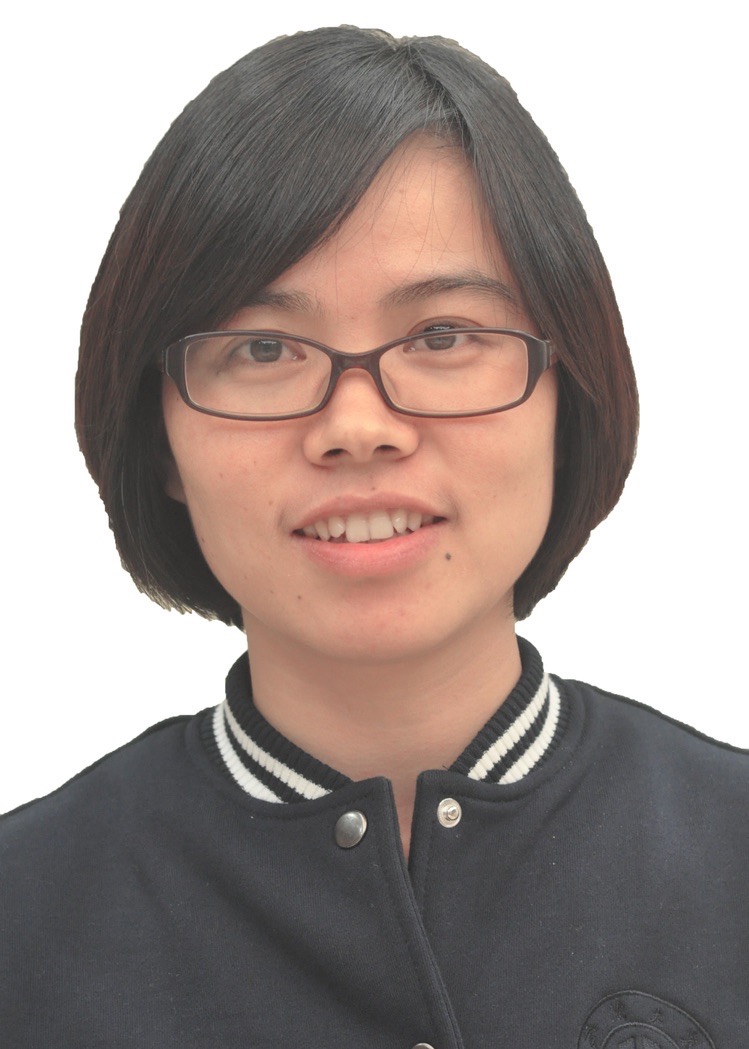}}]{Shuangmin Chen} is a lecturer within School of Information and Technology at Qingdao University of Science and Technology. She got her Ph.D. degree at Ningbo University in 2018, and worked as a research associate in 2009-2012 at Nangyang Technological University (Singapore). Her research interests focus on computer graphics and computational geometry.
She got the Best Paper award of SPM 2017. During the past ten years, she published over 20 research papers on famous journals/conferences.
\end{IEEEbiography}

\begin{IEEEbiography}[{\includegraphics[width=1in,height=1.25in,clip,keepaspectratio]{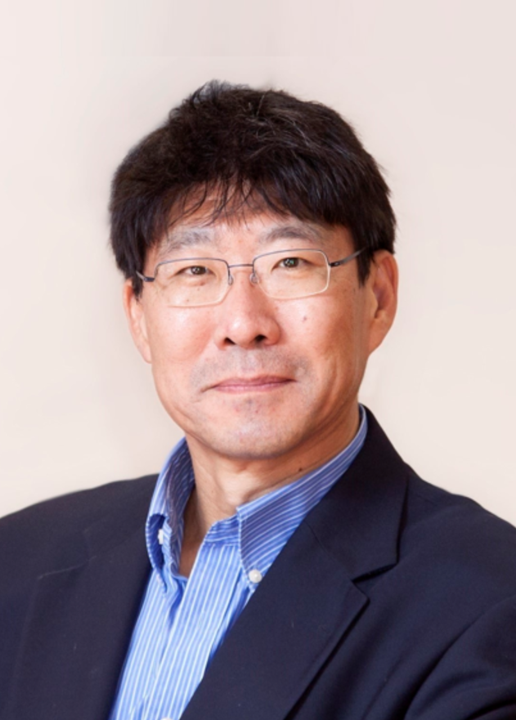}}]{Wenping Wang} obtained his Ph.D. in Computer Science from University of Alberta in 1992. He is Chair Professor of Computer Science at University of Hong Kong. His research interests include computer graphics, computer visualization, computer vision, robotics, medical image processing, and geometric computing. He is associate editor of several premium journals, including Computer Aided Geometric Design (CAGD), Computer Graphics Forum (CGF), IEEE Transactions on Computers, and IEEE Computer Graphics and Applications, and has chaired over 20 international conferences, including Pacific Graphics 2012, ACM Symposium on Physical and Solid Modeling (SPM) 2013, SIGGRAPH Asia 2013, and Geometry Submit 2019. Prof. Wang received the John Gregory Memorial Award for his contributions in geometric modeling. He is an IEEE Fellow.
\end{IEEEbiography}

\begin{IEEEbiography}[{\includegraphics[width=1in,height=1.25in,clip,keepaspectratio]{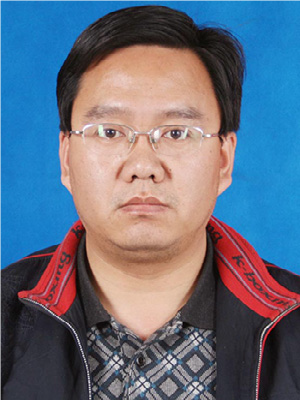}}]{Xiuyang Zhao} is currently a professor in School of Information Science and Engineering, University of Jinan, China. He received Ph.D. in computational material science from Shandong University, China, in 2006. From 2008 to 2010, he held a postdoctoral position in School of Software, Shandong University. His research interests include computer vision, computer graphics and CAGD. He is a member of several academic organizations and a reviewer for many famous international journals.
\end{IEEEbiography}
\vspace{15mm}
\begin{IEEEbiography}[{\includegraphics[width=1in,height=1.25in,clip,keepaspectratio]{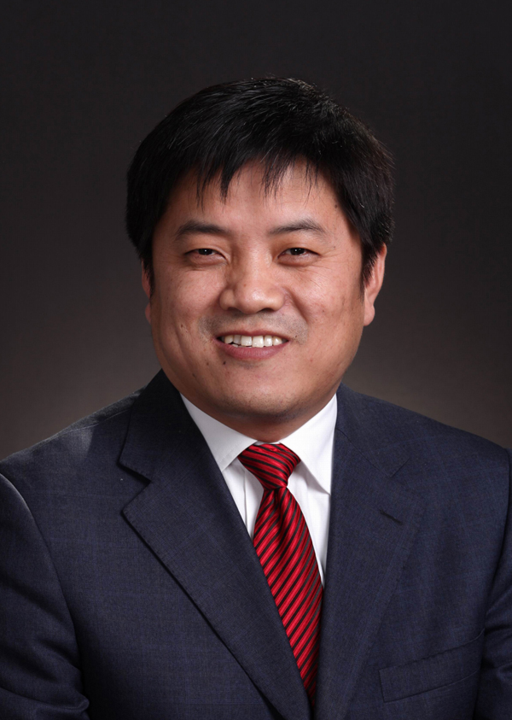}}]{Changhe Tu} received the BSc, MEng, and Ph.D. degrees from Shandong University, China, in 1990, 1993, and 2003, respectively. He is a professor in School of Computer Science and Technology, Shandong University, China. His research interests are in the areas of computer graphics, 3D vision and computer aided geometric design. He has published over 100 papers in international journals and conferences. Currently he leads a CG-VIS group at Shandong University.
\end{IEEEbiography}
\vfill

\end{document}